\documentclass[12pt]{iopart}
\usepackage{float}
\usepackage{bm}
\usepackage{color}
\usepackage{graphicx}
\usepackage[hidelinks]{hyperref}
\usepackage{doi}
\newcommand{\be}{\begin{equation}}
\newcommand{\ee}{\end{equation}}
\newcommand{\benn}{\begin{equation*}}
\newcommand{\eenn}{\end{equation*}}
\newcommand{\bea}{\begin{eqnarray}}
\newcommand{\eea}{\end{eqnarray}}
\newcommand{\ve}{\varepsilon}
\newcommand{\mA}{\mathcal{A}}

\newcommand{\mH}{\mathcal{H}}

\newcommand{\mF}{\mathcal{F}}

\newcommand{\mZ}{\mathcal{Z}}

\newcommand{\ggcol}[1]{{\color{black} #1}}
\newcommand{\myred}[1]{{\color{black} #1}}
\newcommand{\mygreen}[1]{{\color{black} #1}}
\newcommand{\eqref}[1]{(\ref{#1})}

\begin{document}

\title[Localization and classical entanglement in the
  DNLSE]{Localization and ``classical entanglement''\\ in the
  Discrete Non-Linear Schr\"odinger Equation}

\author{Martina Giachello$^{1,2}$, Stefano Iubini$^{3,4}$, Roberto Livi$^{3,4,5}$, Giacomo Gradenigo$^{1,2}$}

\address{$^1$ Gran Sasso Science Institute, Via F. Crispi 7, 67100 L'Aquila, Italy}
\address{$^2$ INFN-LNGS, Via G. Acitelli 22, 67100 Assergi (AQ), Italy}
\address{$^3$ Istituto dei Sistemi Complessi, Consiglio Nazionale delle Ricerche, via Madonna del Piano 10, I-50019 Sesto Fiorentino, Italy}
\address{$^4$ Istituto Nazionale di Fisica Nucleare, Sezione di Firenze, via G. Sansone 1, I-50019 Sesto Fiorentino, Italy}
\address{$^5$ Dipartimento di Fisica e Astronomia and CSDC, Universit\`a degli Studi di Firenze, Italy}

\ead{martina.giachello@gssi.it, stefano.iubini@cnr.it, roberto.livi@unifi.it, giacomo.gradenigo@gssi.it}

\begin{indented}
\item[]March 2025
\end{indented}

\begin{abstract}
We perform a detailed numerical study of the very
peculiar thermodynamic properties of the localized high-energy phase
of the Discrete Non-Linear Schr\"odinger Equation (DNLSE). A numerical
sampling of the microcanonical ensemble done by means of Hamiltonian
dynamics reveals a new and subtle relation between the presence of the
localized phase and \ggcol{a property of the system that we have
called {\it ``classical entanglement''}. Our main finding is that a
quantity defined for our classical system in perfect analogy with
the entanglement entropy of quantum ones, and that we have therefore
called $S_{\mathrm{ent}}$, grows with the system size $N$ in the
localized phase as $S_{\mathrm{ent}}(N) \sim \log(N)$, therefore
revealing the presence of subtle non-local correlations between any
finite portion of the system and the rest of it. This manifestation
of {\it ``classical entanglement''} beautifully captures the lack of
system separability in the DNLSE localized phase, revealing how
statistical correlations specific to the microcanonical ensemble and
non-reproducible in the canonical one, may concur to determine a
property totally analogous to the one produced by non-local quantum
correlations.}
\end{abstract}

%
%
%
%
\newpage
\section{Introduction}
\label{sec:intro}
The dynamic and thermodynamic properties of the high-energy phase of
the Discrete Non-Linear Schr\"odinger Equation (DNLSE), a classical
system that works as an effective description for a bosonic condensate
wavefunction on a lattice~\cite{Kevrekidis09}, have been debated for a
long time. As for high-energy phase, we refer to values of the energy
per lattice site larger than the threshold value $e_{\mathrm{th}} =
2a^2$, where $a$ denotes the mass per lattice site of a bosonic
condensate. According to the seminal results of~\cite{RCKG00}, the
parabola $e_{\mathrm{th}} = 2 a^2$, which is also represented in
Fig.\ref{fig:pd} here denotes the critical line above which the
grand-canonical ensemble becomes ill-defined. Since then, for a couple
of decades, there has been no equilibrium thermodynamic calculation
capable of tackling the regime $e>2a^2$, so that different and
controversial speculations have been made about the nature of this
phase. Numerical simulations of the DNLSE dynamics in this high-energy
regime provided several and incontrovertible evidence that strongly
localized structures (breathers) are stable or, to say the least,
metastable on gigantic timescales which exceed the simulation time
even on modern computers~\cite{Iubini2013_NJP}. Numerical simulations
in the regime $e>2a^2$ provided compelling evidence of a phenomenon
that cannot be described within the grand-canonical ensemble: the
presence of negative temperatures~\cite{Iubini2013_NJP}. We show the
result of numerical simulations which do not suffer this limitation
since they are performed in the microcanonical ensemble, where the
temperature is an observable which can be either positive or negative
since it is directly measured as
\begin{equation}
     \beta = \frac{1}{T}=\frac{\partial S(E)}{\partial E},
     \label{eq:T-micro}
\end{equation}
where $S(E)$ is the microcanonical \ggcol{Boltzmann} entropy as a function of energy
and where we have assumed $k_B=1$. The microcanonical definition of
temperature in Eq.~\ref{eq:T-micro} clearly does not suffer the
limitations that the concept of temperature has in the canonical
ensemble. An account of negative temperatures has been recently
provided in~\cite{Baldovin}: for a system with compact domain
variables, negative temperatures can be consistently defined even in
the canonical ensemble, since the canonical weight $e^{-\beta H}$ is
finite even when $\beta<0$. In contrast, the same is not true for
locally unbounded variables, which is the case for the
DNLSE. Recently, an effective grand canonical description for
delocalized metastable states at negative temperature was provided
in~\cite{IP25}, while a consistent thermodynamic approach capable of
accounting for both localized states and negative temperatures has
been more controversial.\\

By means of a large-deviation exact estimate of the microcanonical
partition function in the regime $e>2a^2$, it has been possible to
provide a complete and precise account of the localized and
negative-temperature phase also from a thermodynamic point of
view~\cite{GILM21a,GILM21b}. Differently from what happens in the case
of long-range interacting systems, where the terminology
\textit{lack of ensembles equivalence} refers to the different
predictions drawn from the two ensembles~\cite{campa09}, for what
concerns the localized phase of the DNLSE, it turns out that a stronger
meaning to the concept of ensemble non-equivalence must be
assumed. The localized phase of the DNLSE {\it only} exists in the
microcanonical ensemble, namely it can be reached only by directly
fixing the energy to a high enough value which cannot be reached in any
way by adjusting the average energy by means of a thermostat
homogeneously coupled to the system. However, this should not be
considered as a surprise or a peculiar situation. It is well-known
from Ruelle's monograph on statistical mechanics~\cite{Ruelle} that,
when localization or condensation phenomena occur, the equivalence
between statistical ensembles may be broken.  In particular, the main
finding of~\cite{GILM21a,GILM21b} was that the localized phase of the
DNLSE is a genuine thermodynamic equilibrium phase characterized by
negative absolute temperatures. The large-deviation calculation of the
microcanonical entropy presented therein was capable of clarifying
first-order features of the localization transition in the DNLSE at
the price of only one relevant, even though standard in the high
energy regime (see also~\cite{arezzo22}), approximation: neglecting
the hopping term in the Hamiltonian.\\

On the basis of the framework just outlined, we can now clearly state
the two main purposes of the present work. First of all, we will
present here a numerical evidence validating the microcanonical
results of~\cite{GILM21a,GILM21b} on the participation ratio and
negative temperature, in particular, we will find a confirmation of the
predicted finite-size scalings. Secondly, and most importantly, we
will show how to characterize ensemble inequivalence, localization and
the breakdown of additivity by measuring a new observable: \ggcol{the
dependence on system size $N$ of a new quantity, which we will define
as the classical analog of entanglement entropy, and which we will
denote in any case (for brevity) as $S_{\mathrm{ent}}(N)$ and refer
to as {\it classical entanglement}.} In the present work, we show how
the condensate wavefunction, which has a quantum origin but is
formally treated as a classical field obeying standard classical
statistical mechanics, in the localized phase, presents features which
are totally analogous to those of a true quantum system: a non-trivial
scaling with system size of \ggcol{$S_{\mathrm{ent}}(N)$, encoding the
presence of {\it ``classical entanglement''}. In particular, we will show, both analytically and numerically, that in the localized phase of the system, which is only captured by the microcanonical ensemble, one has
\begin{equation}
 S_{\mathrm{ent}}(N) \sim \log(N),
\end{equation}
while in the homogeneous phase, where canonical and microcanonical
ensembles are equivalent we find $S_{\mathrm{ent}}\sim \mathrm{const}$. From the physical point of view, this result shows that in the
localized negative-temperature phase, there are subtle global
correlations related to the lack of a thermal ensemble description and
specific of the microcanonical ensemble that prevent the system
separability in the same manner that quantum correlations do for
entangled quantum systems.\\

The presentation of the results in this paper is organized as follows. 
In Sec.~\ref{sec:methods} we introduce the model, recalling the main results 
from~\cite{GILM21a,GILM21b} together with their physical interpretation, which 
are essential for the present study. Sec.~\ref{cteic} describes the simulation 
protocol adopted in this work, with particular emphasis on the choice of initial 
conditions, which plays a crucial role. In Sec.~\ref{sec:therm-obs} we present 
numerical results for the participation ratio $Y_2(N)$ and the microcanonical 
temperature $T_{\mathrm{micro}}(N)$ in the homogeneous and localized phases, 
where their behavior is especially clear. These results are then compared with 
the behavior of {\it ``classical entanglement''}, the central topic of this 
paper, which is discussed in detail in Sec.~\ref{sec:entanglement} through both 
numerical and analytical results. Sec.~\ref{sec:pseudolocalized} is devoted to 
the {\it pseudo-localized} regime, where we again analyze the participation ratio 
and the microcanonical temperature. This regime is treated separately because it 
is pre-asymptotic (disappearing in the thermodynamic limit) and the results are 
more sensitive to the choice of initial conditions, making their thermodynamic 
interpretation more delicate. Finally, Sec.~\ref{sec:conclusions} summarizes our 
conclusions. 

Additional material of a more technical or complementary nature is relegated to the Appendices: the inconsistency of local definitions of temperature in the 
localized phase of the DNLSE is discussed in~\ref{supp:T-nonlocality}; numerical 
indications that the localization transition in the DNLSE displays features of 
an ergodicity-breaking transition (a point deserving further investigation) are 
reported in~\ref{supp:dynamics}; and details on the computation of the 
microcanonical temperature along the Hamiltonian dynamics are given 
in~\ref{supp:F}.}

\section{Localization in the DNLSE: state of the art}
\label{sec:methods}
In this paper, we study numerically the Hamiltonian dynamics of the Discrete Non-Linear Schr\"odinger Equation with two main goals: to confirm the analytical predictions on thermodynamics of~\cite{GILM21a,GILM21b} and to study the behavior of classical entanglement. The DNLSE is a set of classical Hamiltonian equations describing the behavior of a macroscopic field of quantum origin, the condensate wave-function $z_j \in \mathbf{C}$:
\be
i \dot{z_j} = - \frac{\partial \mH}{\partial z_j^*} = -(z_{j+1}+z_{j-1})- 2 |z_j|^2 z_j \,,
\label{eq:eq-motion}
\ee
where the index $j$ labels the sites of a one-dimensional lattice of size $N$ and $i$ is the imaginary constant. Due to global gauge invariance $z_j \rightarrow z_j e^{i\alpha}$ the dynamics of Eq.~\ref{eq:eq-motion} conserves the total mass $\mA$ in addition to the total energy $\mH$, the two quantities reading (in polar representation $z_j = \rho_j e^{i \phi_j}$) respectively as:
\begin{eqnarray}
  \mA &= \sum_{j=1}^N |z_j|^2 \, \equiv \sum_{j=1}^N \, \rho_j^2  \label{norm.1} \\
  \mH &= \sum_{j=1}^N (z_j^*\, z_{j+1} \, + \, z_j\, z_{j+1}^*) \, + \,  \sum_{j=1}^N |z_j|^4 \nonumber \\
  &= 2 \sum_{j=1}^N \rho_j\rho_{j+1} \cos(\phi_{j+1}-\phi_j) \, + \,  \sum_{j=1}^N |\rho_j|^4 
  \label{energy}
\end{eqnarray}
We denote with $E$ and $A$ the real values attributed to the
observables $\mH$ and $\mA$, with $e = E/N$ and $a = A/N$ the
corresponding values per lattice site. Since $A$ and $E$ are the two
conserved quantities that characterize the DNLSE dynamics, it is
natural to retain them as control parameters also for the
thermodynamics and draw a phase diagram in the plane $(a,e)$.\\

\begin{figure}[H]
\centering
  \includegraphics[width=0.7\linewidth]{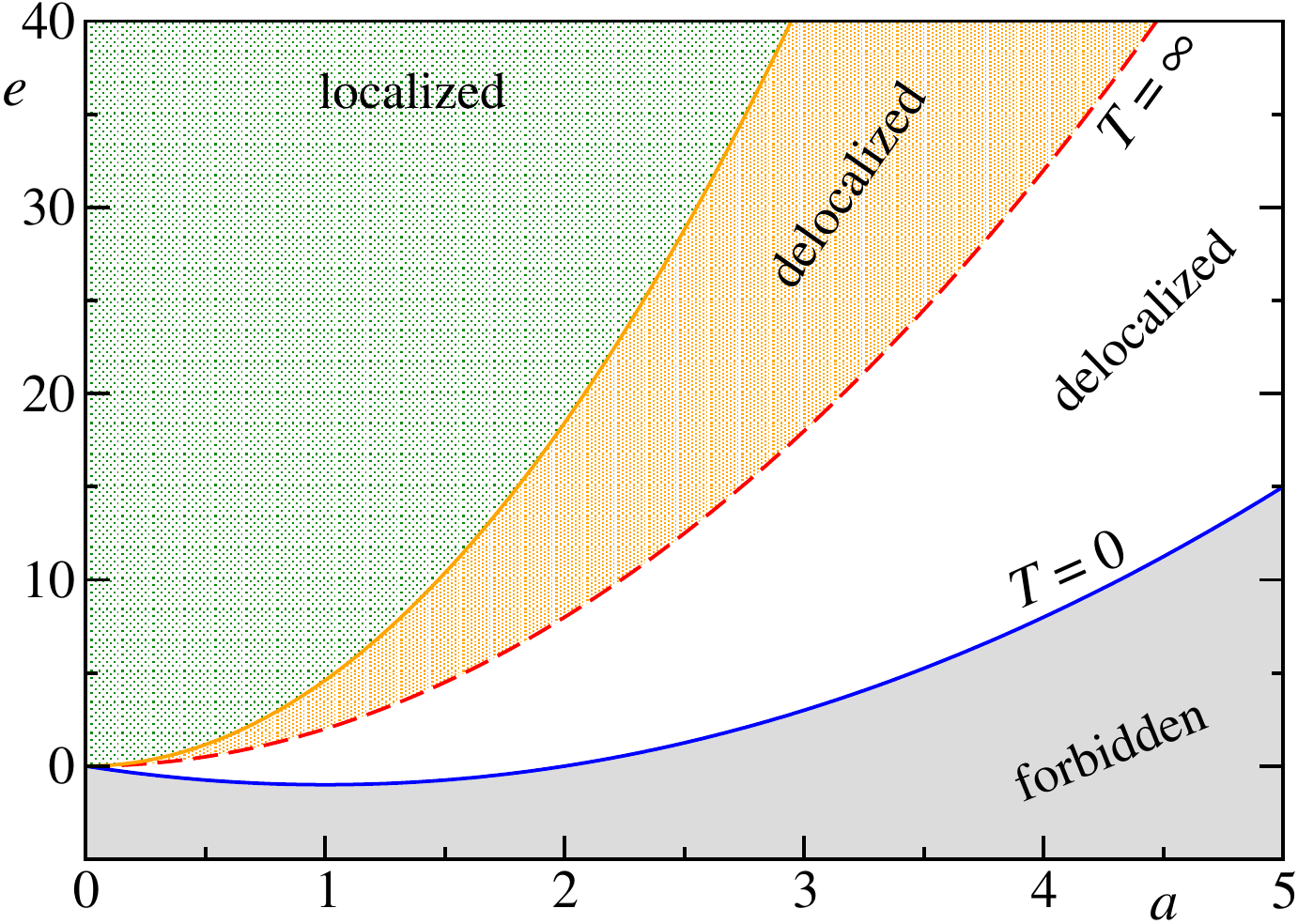}
  \caption{Microcanonical phase diagram of the DNLSE model. Solid blue
    line identifies the ground state $e=a^2-2a$ at $T=0$
    (see~\cite{RCKG00} for details) below which there are forbidden
    states. The red dashed line refers to $e_{\mathrm{th}}(a)=2a^2$
    and is characterized by infinite temperature. Above it $T$ is
    negative: delocalized states are contained in the orange region
    $e_{\mathrm{th}}(a) < e < e_c(a,N)$, while localized ones for $e >
    e_c(a,N)$ are highlighted in green color. The solid orange curve
    here refers to $ e_c(a,N)=e_{\mathrm{th}}(a) + 11.05\, a^2
    N^{-1/3}$ with $N=100$, according to~\cite{GILM21a,GILM21b}.}
  \label{fig:pd}
\end{figure}

\ggcol{The presence of different phases in the DNLSE can be detected, and their properties characterized, through the finite-size scaling of two main observables: the microcanonical temperature $T_{\mathrm{micro}}$, defined as  $T_{\mathrm{micro}}^{-1}=\partial S/\partial E$, and the participation ratio $Y_2$,  which serves as the order parameter of localization. The behavior of these observables in the localized phase was analytically predicted for the first time in~\cite{GILM21a,GILM21b}, where it was shown that only the microcanonical ensemble provides a consistent description.  Before discussing these results and the three phases of the DNLSE, let us briefly recall the key points of the analysis in~\cite{GILM21a,GILM21b}, emphasizing the main outcome of the microcanonical large-deviation estimate developed therein. The central result is that, in order to capture the properties of the localized phase of the DNLSE, it is necessary to compute the system entropy at subleading order in $N$ using a large-deviation approach. This makes it possible to uncover the first-order features of the localization transition: at subleading order in $N$, the entropy clearly reveals the competing contributions of the localized and homogeneous phases. Moreover, only a microcanonical large-deviation calculation allows one to derive analytically the shape of the single-site marginal energy distribution $\mu_E(\varepsilon)$ at total energy $E$ in the localized phase. For other systems exhibiting localization, it was already known~\cite{Majumdar2005_PRL,Szavits2014_PRL} that the transition from the homogeneous to the localized phase can be equivalently described as a change in the behavior of $\mu_E(\varepsilon)$: from a monotonic decay at large $\varepsilon$ to a bimodal distribution with a secondary peak (the condensate bump) at large $\varepsilon$, whose distance from the origin scales as $N$ and whose width scales as $\sqrt{N}$. However, this phenomenon had never been demonstrated for the DNLSE until the results of~\cite{GILM21a,GILM21b}. According to that analysis, at large but finite $N$ the phase diagram of Fig.~\ref{fig:pd} displays three regions above the low-energy forbidden one (gray). Among them, particular importance is carried by the {\it matching regime}, which separates two of the phases and plays a crucial role in understanding the physics of the system. }\\ \\

\begin{itemize}
 \item {\bf Homogeneous phase}: $e<e_{\mathrm{th}}(a)$ \\ The
   distribution of the condensate wave-function on the lattice is
   homogeneous, the temperature is positive, and the ensembles are
   equivalent. \ggcol{That is, the microcanonical temperature
     $T_{\mathrm{micro}}$ is a size-independent observable coinciding
     with the corresponding parameter of the thermal ensemble:
     $T_{\mathrm{micro}} = \beta^{-1}$. In this phase the
     participation ratio vanishes in the thermodynamic limit as
     $Y_2(E) \sim 1/N$ and the marginal distribution
     $\mu_E(\varepsilon)$ decays monotonically as
   \begin{equation} \eqalign{
     \mu_E(\varepsilon) \approx \frac{1}{\sqrt{\varepsilon}} \, \exp{\left( - \sqrt{\varepsilon} \right)}.
   } \end{equation}
   From the above expression it can be easily computed the behaviour of
   the localization transition order parameter, the participation ratio $Y_2(E)$,
   reading as:
   \begin{equation} \eqalign{
     Y_2(E) = \frac{N}{E^2} \int_0^\infty d\varepsilon~\varepsilon^2~\mu_E(\varepsilon) \approx \frac{1}{N}
   } \end{equation}}
   Thermodynamics is standard, i.e., by raising the temperature both
   energy and entropy increase. In this regime, we will show in the
   present work that the \textit{classical entanglement} has a trivial behavior
   which can be traced in any additive system: it does not depend on
   system size.\\ \\ 

 \item {\bf Pseudo-localized phase}: $e_{\mathrm{th}}(a) < e <
   e_c(a,N)$\\ There can be peaks in the condensate wave-function
   distribution on the lattice, but without a concentration of a finite
   fraction of the total energy on a single site. The system can be
   described only in the microcanonical ensemble, temperature is
   negative and thermodynamics is ``non-standard'': the {\it increase}
   of system energy induces a {\it decrease} of entropy, due to an
   increasing tendency to localize rather than share energy. 
   \ggcol{In this regime the local distribution $\mu_E(\varepsilon)$ starts to
     exhibit a non-monotonic decay with a secondary peak, but the
     scaling with $N$ of the peak distance from the origin and of its
     width is not yet the one typical of localization. Indeed, an
     estimate of the participation ratio for energies close enough to
     $e_{\mathrm{th}}(a)$ yields still the same estimate of the
     homogeneous phase, i.e., $Y_2\sim 1/N$~\cite{GILM21a}, while for
     energies closer to $e_c(a,N)$ a precise estimate has not yet been
     provided, although we expect to have $Y_2 \sim 1/N^{\gamma}$,
     with $\gamma \leq 1$.}\\ \\ 
 \item {\bf Localized phase}: $e > e_c(a,N)$ \\ Temperature is
   negative and the system is in the localized phase, where a finite
   fraction of the total energy is concentrated in $\mathcal{O}(1)$
   sites. The ``anomalous'' thermodynamics of the pseudo-localized
   phase is finally established as the dominant behavior: upon
   increasing the energy, the system becomes more and more localized
   and the entropy becomes smaller and smaller. Like a black hole, one
   (or a few) gigantic breather eats all the additional energy which is
   fed to the system, producing a shrinking of phase space and a
   decrease of entropy. We will further clarify the anomalous
   thermodynamic nature of this regime in this work by showing that
   the \textit{classical entanglement} increases logarithmically with the size of the system, indicating the presence of non-trivial global
   correlations and putting on firm conceptual grounds the
   impossibility of providing any {\it local} definition of
   temperature in the system. \ggcol{In this regime is
     possible to give a precise estimate, see~\cite{GILM21a,GILM21b}, of both the behaviour
     of temperature with system size, $T_{\mathrm{micro}} \sim - N^{1/2}$, and of the
     the behavior of the secondary peak of $\mu_E(\varepsilon)$, which is centered at a distance from
     the origin proportional to $N$ and has a width proportional to $N^{1/2}$, allowing to
     write for all practical purposes the marginal energy distribution in this regime as:
     \begin{equation}
       \begin{array}{rcl}
         \mu_E(\varepsilon) &\approx& \frac{1}{\sqrt{\varepsilon}} \, \exp{\left( - \sqrt{\varepsilon} \right)}, \qquad\quad ~\varepsilon \ll \Delta E, \\[8pt]
         \mu_E(\varepsilon) &\approx& \frac{1}{N}\delta\left( \varepsilon - \Delta E \right), \qquad\quad \varepsilon \sim \Delta E~~~\&~~~ N \gg 1.\\ \\ 
       \end{array}
       \label{eq:equilibrium_distribution_0}
     \end{equation}
   .

 \item {\bf Matching regime}: \\
The identification in~\cite{GILM21a} of a so-called {\it matching regime} has been particularly important in revealing the first-order features of the localization transition in the DNLSE. The strategy of the microcanonical calculation, developed for the first time in~\cite{GILM21a}, is to provide large-deviation asymptotic estimates of the microcanonical partition function, with different forms depending on the regime. 
When the total energy $E$ of the DNLSE satisfies $E - E_{\mathrm{th}} \sim N^{2/3}$, where $E_{\mathrm{th}}$ is the total energy at which ensemble equivalence breaks down, it was shown in~\cite{GILM21a} that the partition function reads:
\begin{equation} \eqalign{
  \Omega_N(A,E) = e^{N[1+\log(\pi a)]} 
  \left[ \Omega_N^{\mathrm{cond}}(\zeta) + \Omega_N^{\mathrm{hom}}(\zeta)\right],
  \quad\quad \zeta = \frac{E-E_{\textrm{th}}}{N^{2/3}},
  \label{eq:micro-partition-exact}
} \end{equation}
with
\begin{equation} \eqalign{
  \Omega_N^{\mathrm{cond}}(\zeta) & = \exp\left[ -N^{1/3} \chi(\zeta) \right], \nonumber \\
  \Omega_N^{\mathrm{hom}}(\zeta) & = \exp\left[ -N^{1/3} \frac{\zeta^2}{2\sigma^2(a)} \right], \nonumber \\
  \label{eq:micro-partition-exact-2}
} \end{equation}
where $\Omega_N^{\mathrm{cond}}(\zeta)$ and $\Omega_N^{\mathrm{hom}}(\zeta)$ are respectively the partition functions of the condensed and homogeneous phases. The function $\chi(\zeta)$ has no closed form, but its asymptotic behavior is known: for large $\zeta$, $\chi(\zeta) \sim \sqrt{\zeta}$~\cite{GILM21a}. 
From Eq.~\eqref{eq:micro-partition-exact}, the first-order mechanism of the localization transition becomes transparent. In the limit of large $N$, the system is dominated by the condensed phase when $\chi(\zeta)<\zeta^2/[2\sigma^2(a)]$, and by the homogeneous phase when the opposite inequality holds. This allows one to single out the critical value $\zeta_c$ at which the two phases have equal probability and thus coexist. The analysis of~\cite{GILM21a} finds that $\zeta_c = 2^{1/3} \zeta_l$, where $\zeta_l$ denotes the spinodal point for the existence of the localized phase and depends on the total energy $E$ and the total mass $A$ (see~\cite{GILM21a}). Equation~\eqref{eq:micro-partition-exact} also highlights the crucial subleading contribution to the total entropy of the system, which (with $k_B=1$) in the matching regime takes the form:
\begin{equation} \eqalign{
  S_{N}(A,E) =  
  N \left[ 1 + \log(a\pi) \right] 
  - N^{1/3} \Psi\left( \zeta = \frac{E-E_{\mathrm{th}}}{N^{2/3}}\right),
  \label{eq:entropy-matching}
} \end{equation}
where 
$\Psi(\zeta) = \inf_\zeta \lbrace \chi(\zeta), \zeta^2/(2\sigma^2)\rbrace$. 
Note that in Eq.~\eqref{eq:entropy-matching} the entropy contains an extensive term, which in the localized phase does not depend on the total energy (for $E>E_{\textrm{th}}$) and corresponds to the amount $E_{\mathrm{th}}$ that \textit{always} remains uniformly shared along the lattice. The subextensive term, on the other hand, embodies the physics of the localization transition.
   }
\end{itemize}

In the calculation of the microcanonical partition function, we considered only one approximation: we have neglected the hopping term in Eq.\ref{energy}, thus writing \myred{$\Omega_N$} as
\begin{eqnarray}
  & \Omega_N(A,E) = \int \prod_{i=1}^N dz_i~\delta\left( A - \sum_{i=1}^N |z_i|^2 \right)\delta\left( E - \sum_{i=1}^N |z_i|^4 \right).
  \label{eq:Omega-micro}
\end{eqnarray}
Let us recall the argument which allows to drop the hopping
terms~\cite{GILM21a,GILM21b}: close to and above the
infinite-temperature line $e_{\mathrm{th}}(a)$ one can safely assume
that the phases of the condensate wave-function at different lattice
sites are iid random variables. This implies that the first sum in
Eq.~\ref{energy} is a sum of terms with alternating sign, so that,
while the on-site quartic term of the energy gives a contribution of
order $N$, the hopping term gives a contribution of order
$\sqrt{N}$. Clearly, such an approximation is not rigorously under
control, in particular considering the fact that, from the resulting
simplified partition function, we have computed sub-leading
contributions in $N$ to the microcanonical entropy. This is the reason
why the exact results (under the just-mentioned approximation)
presented in~\cite{GILM21a,GILM21b} need to be validated by numerical
simulations of the full Hamiltonian dynamics where the hopping term is
retained. 

\ggcol{As a final remark, we point out that the analytical estimates here discussed are derived from
the microcanonical partition function in Eq. (\ref{eq:Omega-micro}) and therefore assume also 
a limit of infinite times. 
Recently, a finite-times description  was derived for homogeneous metastable states
lying slightly above $e_{\mathrm{th}}(a)$ in the DNLSE~\cite{IP25}. These states are characterized by a negative
temperature $T$, survive for exponentially long times in $|T|$ and obey a regularized 
grand-canonical statistics. In practice, they correspond to pre-thermal states in which
the localization process has not yet occurred. In this paper we will not explore this metastable regime,
as we will focus on fully developed localized states. Further considerations on metastable states
are contained in Sec.~\ref{sec:pseudolocalized}. 	
}

\section{Methods}
The present study of the DNLSE behavior has two main
goals: confirm the thermodynamic scenario on the localization
transition emerging from the analytic asymptotic estimates
of~\cite{GILM21a,GILM21b} and provide new insights on the system by
studying a new quantity that, in analogy with entanglement entropy for quantum systems, we decided to call \textit{classical entanglement}. To do that, we need the precise knowledge of the energy values at which the system is either in the homogeneous and in the localized phase. From the result of~\cite{GILM21a,GILM21b} we know that the threshold value of the energy $e_{\mathrm{th}}$ at which the description of the system in terms the canonical ensemble breaks down, already found in~\cite{RCKG00}, does not coincide with the critical energy $e_c(N)$ at which localization takes place at large but finite $N$. Therefore, since we have to deal with numerical simulations at finite values of $N$, it is crucial to known $e_c(N)$ in order to choose appropriately the energy where the system is in the localized phase. According to the large deviations estimates of~\cite{GILM21a,GILM21b} these two values have the following relation:
\begin{equation}
e_c(a,N) = e_{\mathrm{th}}(a) + \frac{\zeta_c(a)}{N^{1/3}} \, ,
\label{eq:scaling-ec}
\end{equation}
where both $e_{\mathrm{th}}(a)=2a^2$ and $\zeta_c(a)$ are analytic functions of $a$. The explicit expression of $e_{\mathrm{th}}(a)=2a^2$ has been found both in~\cite{RCKG00} and in~\cite{GILM21a}, whereas the explicit formula of $\zeta_c(a)$ is a result of~\cite{GILM21a}. In particular, for $a=1$ we have
\begin{eqnarray}
  e_{\mathrm{th}}(a=1) &= 2 \nonumber \\
  \zeta_c(a=1) &= 11.05. 
\end{eqnarray} 
Let us notice that while the threshold value $e_{\mathrm{th}}(a) = 2
\,a^2$ can be obtained both from an exact estimate in the
grandcanonical ensemble~\cite{RCKG00} and from the canonical
calculation of~\cite{GILM21a}, the critical value $e_c(a,N)$ where
localization exactly takes place can be obtained only from the
microcanonical calculation. Its precise knowledge is crucial for the
purpose of this paper, allowing a precise choice of initial conditions
either in the homogeneous positive temperature phase or in the
localized phase for any choice of lattice size $N$. The values of
$e_c(N)$ at the different sizes $N$ considered in the present work are
listed in Tab.~\ref{tab:ec-N}. \\

Looking at the phase diagram in Fig.~\ref{fig:pd} it is clear that one
can arbitrarily choose any finite value of $a$ and then study the
properties of the localized phase by increasing $e$. For this reason,
we have fixed the condensate mass to $a=1$ in our numerical simulation
and then varied $e$. Different choices of $a$ are immaterial: \ggcol{the high-energy region is invariant under a uniform rescaling of norms, so that it is quite reasonable to expect equilibrium properties to depend solely on the ratio $e/a^2$.}\\
In the following we will focus on the characterization of three main observables and their dependence on
the lattice size $N$: the {\it participation ratio} $Y_2$, the {\it
  microcanonical temperature} $T^{-1} = \partial S/\partial E$, and
the {\it classical entanglement} $S_{\mathrm{ent}}$.  The crucial point
of our analysis is that, in order to sample the equilibrium value of
these observables by means of the Hamiltonian dynamics, we need to
choose the initial conditions in the corresponding phase. The protocol
followed is described in the next subsection. All numerical
simulations have been performed making use of a symplectic 4th order
integration algorithm of Yoshida type~\cite{yoshida90,boreux10} with
timestep $dt=10^{-2}$, which ensures conservation of energy to order
$dt^4$.

\begin{table}[H]
\centering
\begin{tabular}{|c|c|}   
  \hline $~~e_c(1,N)~~$ & ~~$~~~N~~~$ \\
  \hline 4.19 & 128 \\
  \hline 3.74 & 256\\
  \hline 3.38 & 512 \\
  \hline 3.10 & 1024 \\
  \hline 2.87 & 2048 \\
  \hline 2.69 & 4096 \\
  \hline
\end{tabular}
\caption{Numerical values of the critical energy $e_c(a,N)$ for the localization transition for $a=1$, estimated energy of the {\it``metastable''} phase (negative temperature, no localization) and corresponding $N$.}
\label{tab:ec-N}
\end{table}
%
\subsection{Numerical simulations: Initial conditions}
\label{cteic}

In order to study the localized phase of the DNLSE with Hamiltonian
dynamics we take advantage of the precise knowledge of equilibrium
phases features gathered from~\cite{GILM21a,GILM21b}, which allows us
to choose initial conditions at the appropriate energy $e$ for each
lattice size $N$ in order to be either in the homogeneous or in the
localized phase. This task is particularly delicate in the present
context, since at high energies, due to the lack of a canonical
description, a standard {\it "thermal"} initial condition is
inappropriate and the presence of the localized state must be included
already in the initial state, respecting at the same time the
macroscopic constraints on the condensate total mass and energy. In
particular, when $e>e_{\mathrm{th}}$, the correct procedure is to
assign the initial value of $|z_j|^2=\rho_j$ according to the infinite
temperature canonical distribution on $N-1$ lattice sites and then to
place the remaining mass on the last site. This means that the initial
values of $\rho_j$, for $j=1, \cdots, N-1$ are extracted from the
Poisson-like distribution

\be
 p(\rho_j^2) = \frac{e^{-\rho_j^2/a}}{a}
\label{eq:distr-poiss} 
\ee
while a ``breather" containing the excess energy is placed on a randomly chosen site on the lattice, for which we considered periodic boundary conditions. On this site the value of the density is given by the expression
\be
  \rho_N^2 = \sqrt{e - e_{\mathrm{th}}(a)}~N^{1/2} \, .
\ee
The phases $\phi_j$ are randomly initialized from a uniform probability distribution in the interval $[ 0 , 2 \, \pi ]$. The values of $ \rho_j^2$, initially sampled from the distribution in Eq.~\ref{eq:distr-poiss}, are finally rescaled in order to fulfill the macroscopic constraints on mass and energy:
\begin{eqnarray}
  A = a~N = \sum_{j=1}^N \rho_j^2& \qquad\qquad E = e~N = \sum_{j=1}^N \rho_j^4 
  \label{eq:constraints}
\end{eqnarray}
This sort of initial conditions is obviously stable with respect to Hamiltonian dynamics in the localized phase. In the positive temperature phase, on the other hand, the system is initialized according to the following procedure: first, similarly to the localized phase, we initialize the system at infinite temperature, following the distribution of Eq.~\ref{eq:distr-poiss}. Then, using a phase update algorithm, we progressively decrease the energy down to the desired value $e<e_{\mathrm{th}}$. The phase update algorithm works by selecting a random site on the lattice and changing its phase by drawing a new value from $[-\pi,\pi]$. If the new phase decreases the energy density, it is accepted; otherwise, it remains unchanged. This procedure is repeated until the desired energy density $e$ is reached.
\section{Participation ratio and microcanonical temperature}
\label{sec:therm-obs}
In this section it is presented the numerical validation of the main thermodynamic results from~\cite{GILM21a} on the participation ratio and the finite-size scaling of the DNLSE microcanonical temperature in the localized phase. In particular we focus here on the homogeneous and on the fully localized regimes, corresponding respectively to $e<e_{\mathrm{th}}$ and $e>e_c$, leaving the discussion of the pseudo-localized phase at $e_{\mathrm{th}} < e < e_c$ in Sec.~\ref{sec:pseudolocalized}.
\newpage
\begin{figure}[H]
    \centering
\includegraphics[width=0.75\linewidth]{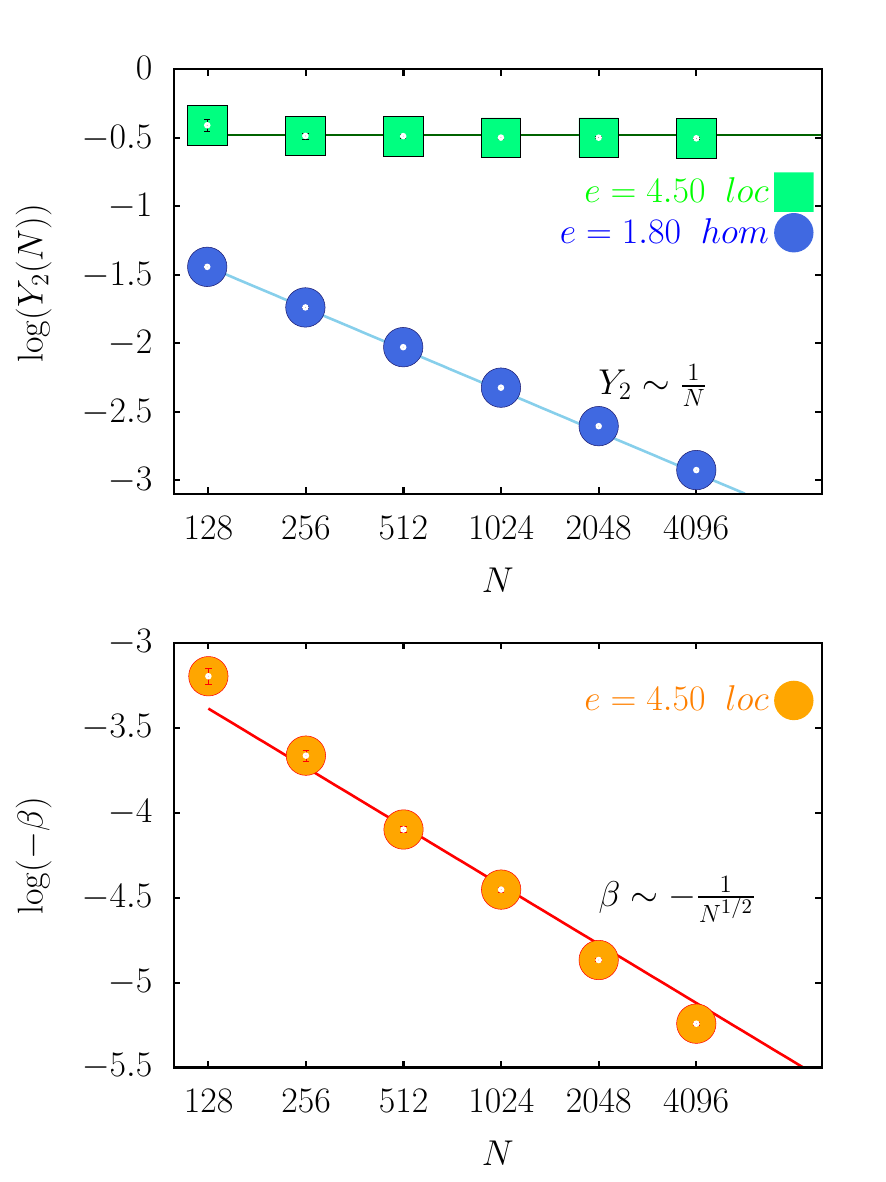}
    \caption{$\mathrm{Upper panel:}$ Finite-size scaling of the participation ratio $Y_2(N)$ of the DNLSE for $a=1$ and different energy densities $e$. Statistical expectations are computed as time averages in the time interval $10^5$-$10^6$. Error bars are the standard deviation of fluctuations, sampled at intervals $\Delta t =10$, in the measurement window. $\mathrm{Lower panel:}$ Finite-size scaling of minus the inverse temperature, $-\beta$, vs. $N$ in the metastable and stable localized phases. The system has been initialized with an artificial breather added to an infinite temperature background, with $a = 1$. The running time of the simulations is $\Delta \tau = 1.5 \times 10^3$, and the microcanonical temperature has been measured globally across all the sites of the system. The uncertainties have been evaluated as the standard deviation of the temperature measurements taken at every $d\tau = 10$ as the simulations progress. The red line corresponds to the fit with the theoretical prediction.}
    \label{fig:part_rat_temp_observables}
\end{figure}
\subsection{Participation Ratio}
\label{subsPR}
The order parameter which signals the presence of a localized phase is the participation ratio $Y_2(N)$, which is defined as follows:
\begin{equation}
Y_2(N) = \left\langle \frac{\sum_{j=1}^N \ve_j^2}{\left(\sum_{j=1}^N \ve_j\right)^2}\right\rangle_{A,E},
\label{eq:part-ratio}
\end{equation}
where $\left\langle ~\right\rangle_{A,E}$ denotes the microcanonical average and we have introduced the shorthand notation $\varepsilon_j = \rho_j^4$. The analytic calculation reported in~\cite{GILM21a} provides precise estimates about the dependence of $Y_2(N)$ on the system size $N$ in two regimes, the localized phase at $e > e_c$ and the the homogeneous phase at $e < e_{\mathrm{th}}$. For intermediate values of the energy, $e_{\mathrm{th}} < e < e_c$, a solid argument is offered in~\cite{GILM21a} only for differences of order $e-e_{\mathrm{th}} \sim N^{-1/2}$, but not for differences in the so-called {\it matching} regime, $e-e_{\mathrm{th}} \sim N^{-1/3}$, where one can only guess a scaling of the kind $e-e_{\mathrm{th}} \sim N^{-\gamma}$, with $\gamma < 1$.\\
It is also worth recalling that this intermediate regime shrinks to zero in the thermodynamic limit, since $\lim_{N\rightarrow\infty} e_c(N) = e_{\mathrm{th}}$. \\
In the upper panel of Fig.~\ref{fig:part_rat_temp_observables} is reported the dependence on $N$ of $Y_2 (N)$ for the two energy regimes just mentioned, obtained by fixing the condensate mass per lattice site to $a=1$. For $e=1.8$ (blue circles) the system is in the positive temperature region (recall that $e_{\mathrm{th}} =2$) and numerical simulations confirm the reliability of the analytic estimate, i.e. $Y_2(N)$ is actually found to decay as $N^{-1}$, as is emphasized by the straight blue line. For the localized phase at $e=4.5 > e_c$ (green squares) there is also a compelling evidence of the agreement between the simulations of Hamiltonian dynamic and the analytic estimates of~\cite{GILM21a,GILM21b}, i.e., $Y_2(N)$ approaches a constant value $Y_2(\infty)$ for large $N$, as is emphasized by the straight green line, which is compatible with asymptotic estimate $Y_2(\infty)=(e-e_{th})^2/e^2$~\cite{GIP21}. The validation of the theoretical predictions presented here and obtained from the full Hamilton dynamics is not redundant with respect to the agreement between theory and numerics already found in~\cite{GIP21,GIP24} by means of stochastic algorithms, since also in this last case the hopping terms were neglected.\\
\subsection{Negative Temperature}
\label{sec:subTneg}
Here we present evidence from numerical data that in the localized phase, $e>e_c(N)$, there is an agreement between the measure of negative temperature along Hamiltonian dynamics and the analytical estimate of~\cite{GILM21a,GILM21b}. Let us stress that, in the microcanonical ensemble, temperature is an observable, not a parameter. For the system considered here, two issues arise in its measurement: one related to the specific Hamiltonian defining the DNLSE and another to the phase of interest. The first issue stems from the $\mathrm{non-separability}$ of the DNLSE Hamiltonian~(\ref{energy})~\cite{franzosi11,iubini2012,levy18,levy21}, as its canonical conjugate momentum is proportional to \( z_j^* \) rather than \( \dot{z}_j \). It is therefore not possible by construction to define operatively temperature as the average kinetic energy of the systems, as it is usually done in molecular dynamics simulations, due to the non-separability of the Hamiltonian. The only definition of temperature which can be exploited in this context is the microcanonical definition: $T^{-1}= \partial S/ \partial E|_A$, where $S$ is the total entropy~\cite{franzosi11}. As reported in the literature on the subject, it is possibile to perform a direct measurement of the microcanonical temperature by considering an Hamiltonian observable defined as a {\it global} function $\mathcal{F}\{z_j\}$ of all variables $z_j$~\cite{rugh97,franzosi11,Baldovin}. The microcanonical temperature is obtained as $T^{-1} =\langle \mathcal{F}\{z_j\}\rangle_E$, where the brackets $\langle \cdot \rangle_E$ denote expectation in the microcanonical ensemble at energy $E$, which in the numerical simulation is realized as time average along the Hamiltonian dynamics with initial conditions at $E$. Details on the expression of $\mathcal{F}\{z_j\}$ are provided in~\ref{supp:F}.\\
By exploiting the above definition, temperature has been measured as a global observable in numerical simulations at $e>e_c(N)$, starting from localized initial conditions. In the second panel of Fig.~\ref{fig:part_rat_temp_observables} it is shown the perfect agreement with the analytical predictions of~\cite{GILM21a,GILM21b} from the numerical study of the Hamiltonian: the temperature is negative and grows with system size as
\begin{equation}
 T \sim - N^{1/2}.
\end{equation}
We leave to~\ref{supp:T-nonlocality} the comment on the behavior of temperature in the pseudo-localized energy regime, $e_{\mathrm{th}}< e < e_c(N)$. Of particular importance is to notice that in the localized regime, $e>e_c(N)$, the only correct definition of temperature is the global one: any local measurement, that is any measure obtained by computing $\mathcal{F}\{z_j\}$ only on a subset of lattice sites, provide a wrong result.\\
At this point any skeptical reader may object that there is freedom in the choice of temperature definition, as long as it is in agreement with the common physical intuition. Why should the global definition be the only correct one? Why can we not say that all chunks of the DNLSE lattice with a homogeneous distribution of energy have their own temperature, which is infinite? The answer to this question comes from the results on classical entanglement. In the localized phase, the system is not separable; any small portion remains entangled with the rest of the system even in the limit $N\rightarrow\infty$, a sort of correlations which have been so far dubbed of purely quantum origin. As will be thoroughly discussed in the next section, we have discovered that a non-trivial behavior of entanglement can also be found for a classical system in those phases, like the localized phase of the DNLSE, which are induced by the presence of an ineludible global constraint. 
\ggcol{\section{Classical entanglement}}
\label{sec:entanglement}
\ggcol{In this section, we show how the characterization of the system non-separability in the localized phase of the DNLSE, which rules out any possibility of a local definition of temperature, is elegantly captured by the behaviour of classical entanglement, which we are going to introduce. Usually, a non-trivial behaviour of entanglement entropy is considered a purely quantum property~\cite{Henderson_2001}, but an analogous quantity, which precisely like entanglement entropy captures subtle non-local correlations, can be defined in full generality also for classical systems. To introduce this observable for the DNLSE, it is convenient to recall the general definition of entanglement entropy in quantum field theory, see, e.g.~\cite{CH09}. Before giving the formal definition, it is worth recalling the physical meaning of the entanglement entropy of a portion $A$ of any given quantum system, in contrast with the Boltzmann entropy of the same subsystem $A$. While the Boltzmann entropy of $A$ carries information exclusively on the degrees of freedom contained in $A$, and more precisely {\it ``counts''} the number of microstates available $\mathcal{N}_A$ through the formula $S(A) = k_B \log(\mathcal{N}_A)$, the entanglement entropy carries information about the rest of the system, which we denote as $B = V \setminus A$, if $V$ is the whole system. Let the total number of degrees of freedom be $N=N_A+N_B$. Clearly, while the Boltzmann entropy $S(A)$ of the subsystem $A$ will always be a function only of $N_A$, the specific property of entanglement entropy is to retain a dependence on $N$ when $N_A$ is kept fixed and $N_B$ is increased, provided that non-local correlations between $A$ and $B=V\setminus A$ are present. The formal definition is as follows. Let us consider a quantum field $\phi$, without even specifying whether the theory is interacting or not. In full generality the state of the system can be specified by a wave-{\it functional} $\Psi(\phi)$, which obeys the {\it functional} Schr\"odinger equation $i\hbar \frac{d}{dt}|\Psi(\phi)\rangle = \hat{\mathcal{H}}(\phi) |\Psi(\phi)\rangle$, see, e.g.~\cite{Hatfield:1992rz}. If we define a sub-portion $A$ of the space-time volume $S$ where the field is different from zero, the entanglement entropy of $A$ is obtained by first computing the {\it functional} reduced density matrix,
\begin{equation} \eqalign{
  \hat{\boldsymbol{\varrho}}_A(\phi) &= \textrm{Tr}_{S\setminus A}|\Psi(\phi)\rangle \langle \Psi(\phi) | \label{eq:entanglement-QFT} \\
  S_A^{\textrm{ent}} &= - \textrm{Tr}[\hat{\boldsymbol{\varrho}}_A(\phi) \log \hat{\boldsymbol{\varrho}}_A(\phi)] \label{eq:entanglement-QFT-2}
} \end{equation}
When, for a fixed subset $A$, the entropy $S_A^{\textrm{ent}}$ depends on the total measure of $S$, we say that $A$ is entangled with the rest of the system, i.e., it has non-trivial quantum correlations such that $\hat{\boldsymbol{\varrho}}_A(\phi)$ retains information about arbitrarily distant regions of space-time. Now let us see how a classical analog of this quantity can be defined, capturing essentially the same information—what is usually called {\it entanglement} for quantum systems and {\it non-separability} for classical systems. For simplicity, we may imagine that the sets $A$ and $S$ are sites of a lattice, and that $\phi$ is the non-relativistic Schr\"odinger field, for which entanglement entropy is in principle obtained from Eqs.~\eqref{eq:entanglement-QFT},~\eqref{eq:entanglement-QFT-2}. In certain situations, such as in the presence of a large Bose-Einstein condensate fraction, the field $\phi$ behaves ``classically'', i.e., most bosons are described by the {\it condensate wave-function} $\langle \phi_i \rangle = z_i$. Since the dynamics of $z_i$ is classical Hamiltonian, we can then, under reasonable ergodicity assumptions, claim that it has an equilibrium microcanonical distribution $\varrho(z_1,\ldots,z_N)$, and replace the functional density matrix accordingly:
\begin{equation} \eqalign{
  |\Psi(\phi_i)\rangle \langle \Psi(\phi_i) | ~~\longrightarrow~~ \varrho(z_1,\ldots,z_N)
} \end{equation}
so that quantum fluctuations are replaced by classical fluctuations associated with the microcanonical distribution. The reduced functional density matrix of the quantum field is then replaced with the marginal distribution:
\begin{equation} \eqalign{
  \varrho_A &= \textrm{Tr}_{\lbrace z_j | j \in S\setminus A \rbrace }\left[ \varrho(z_1,\ldots,z_N) \right] \label{eq:entanglement-classical} \\
  S_A^{\textrm{ent}} &= - \textrm{Tr}[\varrho_A \log \varrho_A ] \label{eq:entanglement-classical-2}
} \end{equation}
One can then say that the system has {\it ``classical entanglement''}, meaning that it is not additive, if by increasing the size of $V$ while keeping $A$ fixed, the entropy $S_A^{\textrm{ent}}$ retains a dependence on $V$. In this spirit, {\it ``classical entanglement''} encodes precisely the same physical information as quantum entanglement: the lack of measure factorization for distant subsets of the measure used to compute physical expectation values. Hence, the commonality between the {\it ``classical entanglement''} discussed here and the well-known quantum entanglement discussed throughout the literature (see, e.g.,~\cite{CC04, Alba:2017ekd, Alba:2017lvc}) becomes manifest. The two notions of entanglement express the lack of system additivity, or, equivalently, the impossibility of separating the system into statistically independent subsystems. The difference between them lies in the origin of correlations: in the classical case, correlations are intrinsic to the microcanonical ensemble, especially in regimes where this ensemble has no thermal counterpart, whereas in the quantum case they are a consequence of the correlations specific to quantum systems. Another distinction is that, while in most works on quantum integrable systems the {\it spreading} of entanglement is explained in terms of the correlated dynamics of quasi-particle modes, in the present discussion, the analysis is purely thermodynamical. The system considered here is not integrable, and no reference is made to correlated dynamics of long-lived excitations related to integrability or quasi-integrability.  

To better grasp the meaning of the quantity $S_A^{\textrm{ent}}$ defined in Eq.~\eqref{eq:entanglement-classical-2}, it is useful to compare it with the Boltzmann entropy. The latter is simply defined as the total volume of phase space available to the system under the microcanonical constraint ($k_B=1$):
\begin{equation} \eqalign{
  S(E) = \log \left(\int \prod_{i=1}^N dz_i~\delta \!\left[ E - \mathcal{H}(z_i) \right]\right),
} \end{equation}
and it always refers to the {\it whole} system. By contrast, the classical entanglement captured by $S_A^{\textrm{ent}}$ quantifies how strongly the subset $A$ is correlated with the rest of the system. Restricting the Boltzmann entropy to the subsystem $A$ is straightforward only when the probability distribution of the degrees of freedom in $A$ factorizes with that of the rest; in such a case $S_A^{\textrm{ent}}$ may coincide with the Boltzmann entropy of $A$. Otherwise, when factorization fails, the definition of $S_A^{\textrm{ent}}$ remains well posed, while the Boltzmann entropy of $A$ is ill defined in the microcanonical ensemble, since the subsystem is not isolated.} Because there is no restriction on the choice of $A$, one may take $A$ to consist of a single degree of freedom, e.g.\ $A=z_j$ with arbitrary index $j$.  
\newpage
\noindent
The next step is to compute the marginal probability distribution $\mu_A$ of the subsystem $A$:
\begin{equation}
  \mu_A = \mu(z_j) = \mathrm{Tr}_{B}\big[\varrho(z_1,\ldots,z_N)\big],
\end{equation}
where, in the quantum formalism, the role of $\mu_A$ is played by the reduced density matrix $\hat{\boldsymbol{\varrho}}_A$. The entanglement is then given by the Von Neumann entropy of $\mu_A$ (or of the reduced density matrix in the quantum case):
\begin{equation} \eqalign{
  S_{\mathrm{ent}} = - \mathrm{Tr}[\mu_A \log \mu_A] 
  = - \int dz_j \,\mu(z_j)\,\log\mu(z_j).
} \end{equation}

Let us note that the above definition of single-site classical entanglement in the DNLSE is formally analogous to the \textit{particle entanglement entropy} discussed in~\cite{SSC07,HZS09}. Due to the arbitrariness in the choice of $A$ and $B$, which can be made simply for convenience, the notation $S_{\mathrm{ent}}$ does not usually carry a label tracking the subsystem choice. The information content of entanglement entropy is the following: if, even in the thermodynamic limit and away from a critical point, subsystem $A$ remains correlated with arbitrarily distant portions of the system $S$, then $S_{\mathrm{ent}}$ retains a dependence on the system size $N$. Otherwise, $S_{\mathrm{ent}}$ does not depend on $N$. Typically, in classical systems with many degrees of freedom and away from criticality, no such correlations exist, which explains why a non-trivial behaviour of entanglement entropy has long been considered a purely quantum property with no classical analog. The case considered here provides the first example of a classical phase where such correlations can be detected and a classical explanation of their origin can be given. In the localized phase of the DNLSE, entanglement arises precisely because this is a purely microcanonical phase with no thermal analog, due to the lack of factorization in the joint distribution $\varrho(z_1,\ldots,z_N)$.  

\begin{figure}[H]
  \centering
  \includegraphics[width=0.75\linewidth]{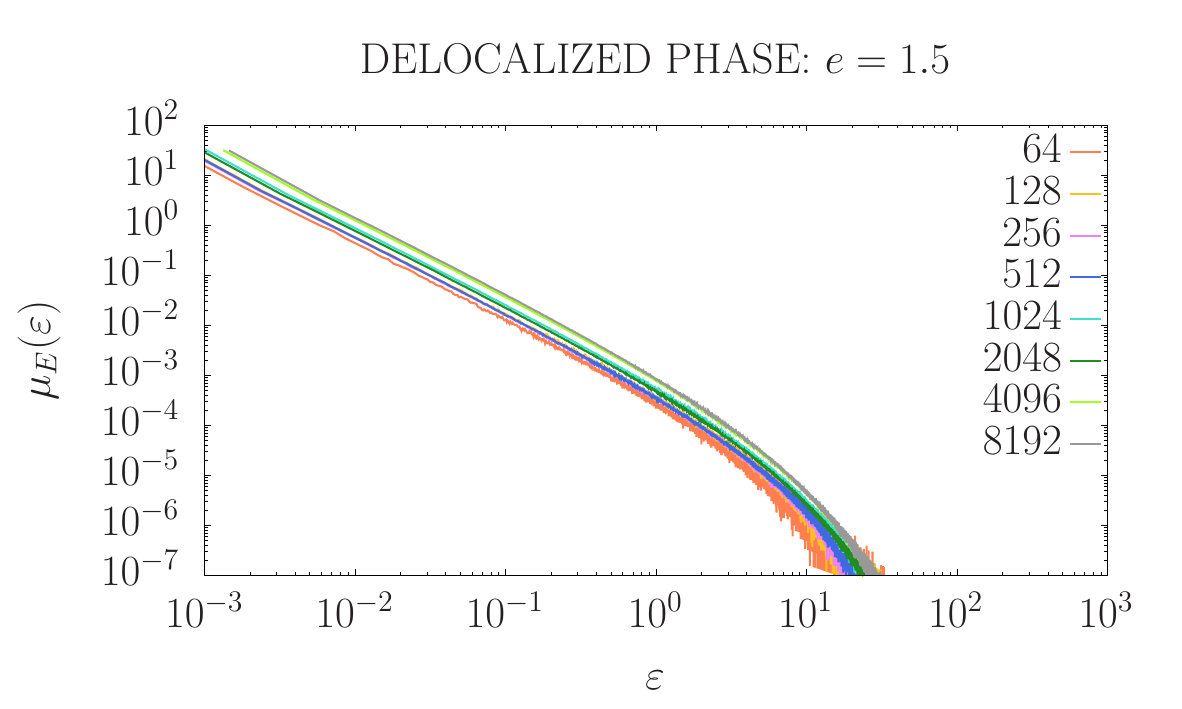}
  \caption{Marginal probability distribution of the local energy $\varepsilon_j=|z_j|^4$ for a single site of the lattice, averaged over all sites. The system is in the homogeneous positive temperature phase: $E/N = 1.5 < e_{\mathrm{th}}$.}
  \label{fig:marg-hom}
\end{figure}

\begin{figure}[H]
  \centering
  \includegraphics[width=0.75\linewidth]{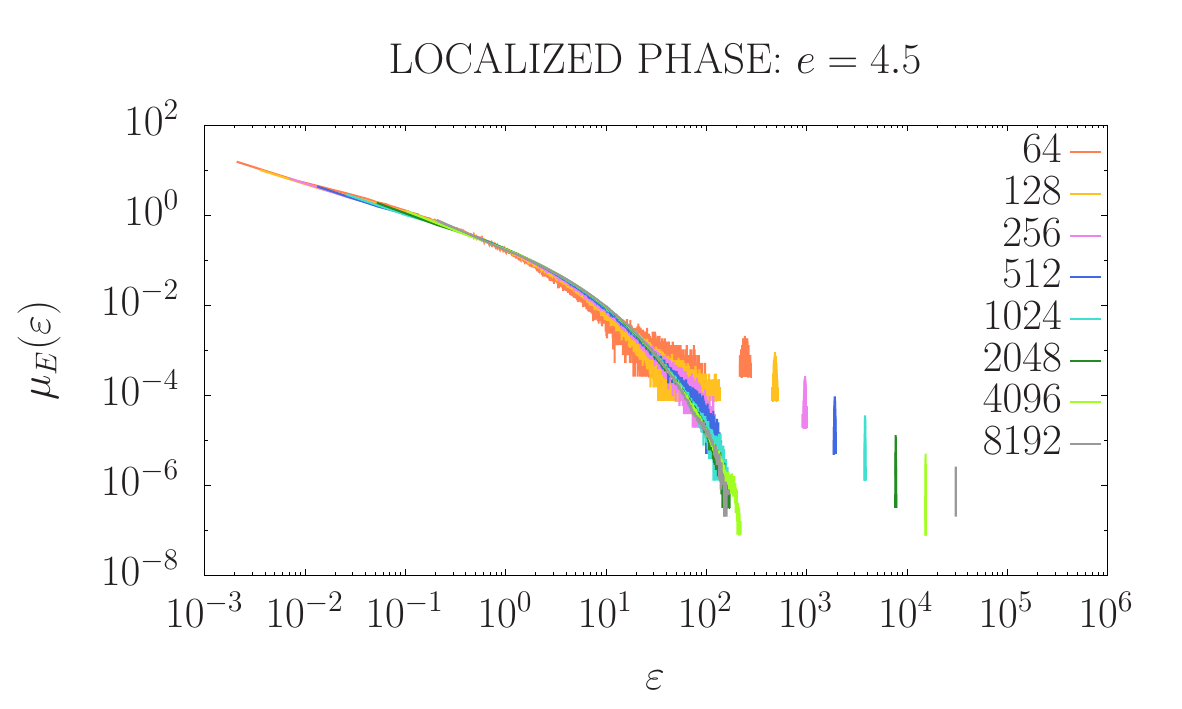}
  \caption{Marginal probability distribution of the local energy $\varepsilon_j=|z_j|^4$ for a single site of the lattice, averaged over all sites. The system is in the localized negative temperature phase: $E/N = 4.5 > e_c$. Secondary peaks represent the characteristic condensate bump.}
  \label{fig:marg-loc}
\end{figure}
\begin{figure}[H]
\centering
\includegraphics[width=0.75\linewidth]{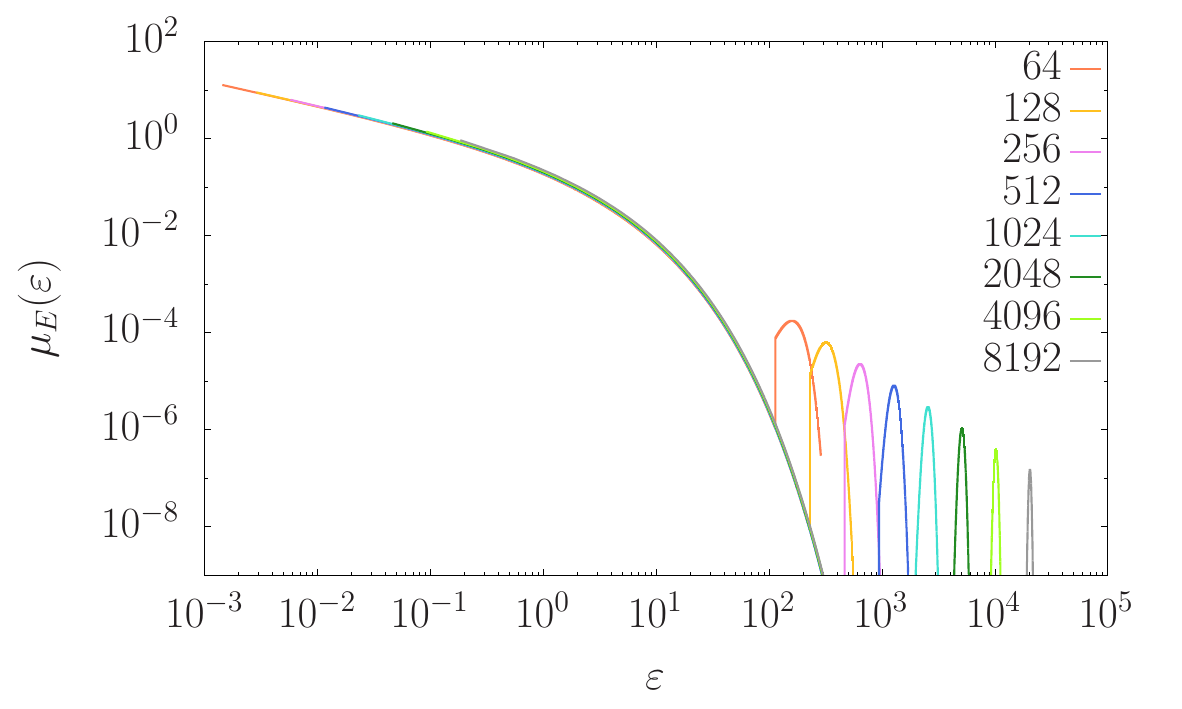}
\caption{Analytical result for the marginal distribution $\mu_E(\varepsilon)$ at different values of the parameter $N$, see Eq.~\ref{eq:marg-th}. The total energy is fixed to $E = N e$, with $e=4.5 > e_c$: the system is in the localized state.}
\label{fig:marg-th}
\end{figure}
\begin{figure}[H]
\centering
\includegraphics[width=0.75\linewidth]{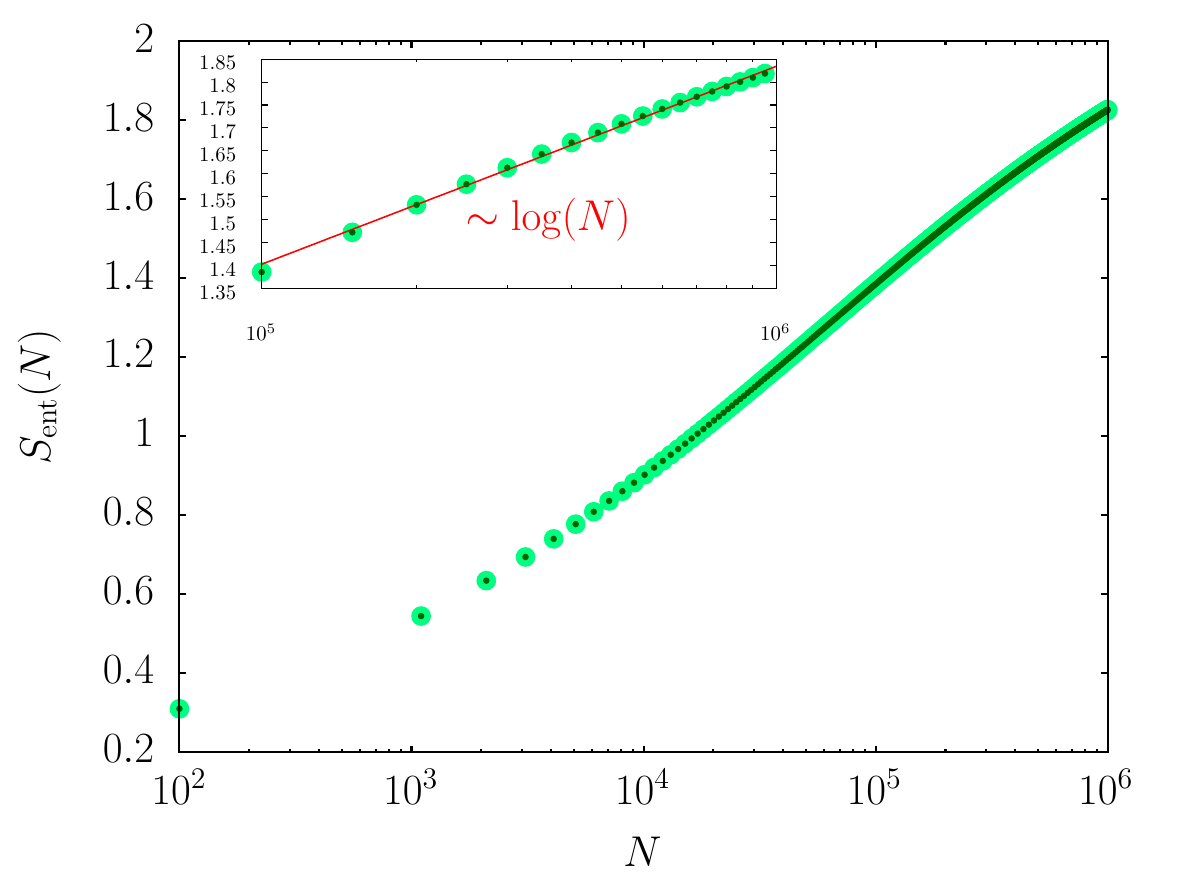}
\caption{Classical entanglement computed from numerical integration of $\mu_E(\varepsilon)\log\mu_E(\varepsilon)$, where the expression of $\mu_E(\varepsilon)$ is the one of Eq.~\ref{eq:marg-th}.}
\label{fig:ent-th}
\end{figure}

\ggcol{\subsection{Classical entanglement: Numerical Results}}
\noindent
Here are presented the numerical results for the measure of classical entanglement in the delocalized and localized phases of the DNLSE. Since localization takes place for energy, we consider for the measure of entanglement the local energies $\varepsilon_j =|z_j|^4$. We are therefore led to consider the joint probability distribution $\varrho_E(\varepsilon_1, \ldots, \varepsilon_N)$, whose exact analytical expression is provided in Eq.~(59) of~\cite{GILM21a}. The subscript $E$ explicitly indicates that the properties and shape of this distribution depend on the total energy $E$. We then define the marginal distribution for a single site of the DNLSE lattice:
\begin{equation}
\mu_E(\varepsilon_j) = \mathrm{Tr}_{\lbrace \varepsilon_i \neq \varepsilon_j \rbrace}[\varrho_E(\varepsilon_1, \ldots,\varepsilon_N)].
\label{eq:ent-marginal}
\end{equation}
Since the choice of $\varepsilon_j$ is arbitrary, given that the
lattice is homogeneous and we have periodic boundary conditions, we
can drop the site index in the definition of
Eq.~\ref{eq:ent-marginal}: $\mu_E(\varepsilon_j) =
\mu_E(\varepsilon)$. \mygreen{It could be argued that a localized
  breather breaks the homogeneity and that the above argument is
  wrong: this is not the case because the marginal distribution
  $\mu_E(\varepsilon)$ must be computed with respect to the whole
  microcanonical ensemble, therefore including all the possible
  locations of the breather on the lattice. Therefore, while any
  specific choice of an initial condition breaks homogeneity along the
  dynamics, at the same time from the statistical point of view the
  system is always homogeneous, roughly speaking because all localized
  configurations have the same probability to be sampled, and the
  numerical procedure to compute both $\mu_E(\varepsilon)$ and the
  classical entanglement must take this into account.} \myred{In other
  words, the precise location of the breather on the lattice is only
  pertinent to the choice of initial conditions in the dynamics, with
  respect to which the equilibrium thermodynamics must be agnostic. It
  is precisely for this reason that the correct definition of the
  marginal distribution $\mu_E(\varepsilon)$, given here below in
  Eq.~\eqref{eq:mue-numeric}, must be given as an average over all
  sites of the lattice: this averaging procedure morally corresponds
  to the average over all possible initial conditions on the
  microcanonical shell.} While for the analytical estimate of
$\mu_E(\varepsilon)$ we will use the formulae of~\cite{GILM21a},
obtained from the large-deviation estimate of the microcanonical
entropy, its numerical sampling is much simpler. Assuming the
Hamiltonian dynamics of the DNLSE to be our importance sampling
protocol for the microcanonical ensemble, the numerical estimate of
$\mu_E(\varepsilon)$ is simply obtained by computing empirically
$\mu_E(\varepsilon_j)$ from the fluctuations along the dynamics and
then averaging:
\begin{equation}
  \mu_E(\varepsilon) = \frac{1}{N}\sum_{j=1}^N \mu_E(\varepsilon_j)
  \label{eq:mue-numeric}
\end{equation}
\begin{figure}[H]
\centering
  \includegraphics[width=0.75\columnwidth]{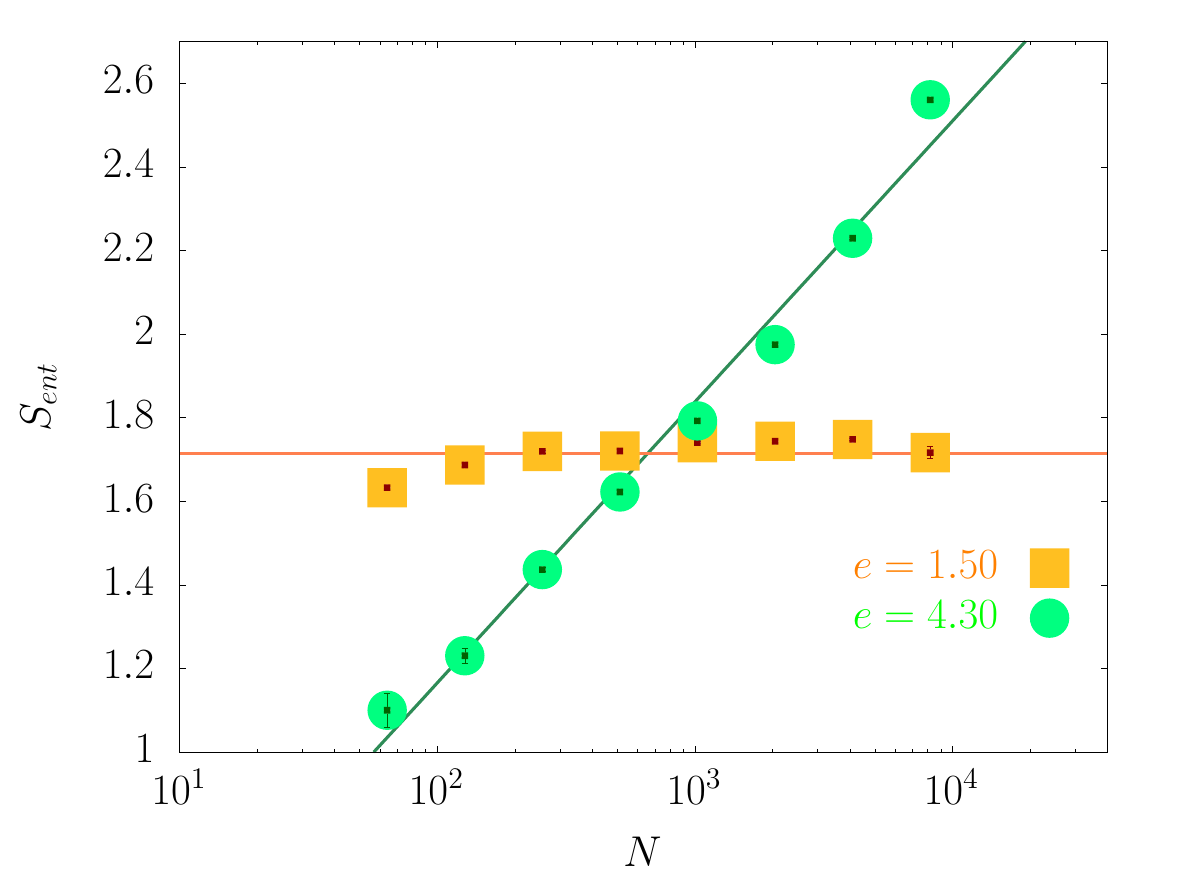}
  \caption{Behavior of the classical entanglement $S_{\mathrm{ent}}(N)$ measured numerically as a function of lattice size $N$. {\it Orange squares}: classical entanglement of the homogeneous phase at positive temperature, $(e = 1.5) < e_{th}$: {\it Green circles}: localized phase at negative temperature, $(e = 4.5) > e_c$. In the homogeneous phase $S_{\mathrm{ent}}(N)$ is constant
    while in the localized phase it grows as $S_{\mathrm{ent}}\sim
    \log(N)$, as emphasized by the straight line in the semi-log
    scale.}
  \label{fig:entg-entropy-num}
\end{figure}
In Fig.~\ref{fig:marg-hom} and Fig.~\ref{fig:marg-loc} is shown the shape of $\mu_E(\varepsilon)$ computed numerically at different lattice sizes $N$, respectively for the homogeneous and the localized phase. In particular, in Fig.~\ref{fig:marg-hom} the marginal distribution $\mu_E(\varepsilon)$ is computed for $E/N = 1.5 <e_{\mathrm{th}}$, where the temperature is positive and localization is absent. In Fig.~\ref{fig:marg-loc} are shown marginals sampled at $E/N = 4.5 > \varepsilon_c$, where temperature is negative and the distributions exhibit the characteristic condensate bump~\cite{Majumdar2005_PRL,GILM21a,Mori_2021}. From the numerically sampled marginals we then obtained the numerical estimate of classical entanglement by simply applying the definition:
\begin{equation}
S_{\mathrm{ent}}(N) = - \sum_{\varepsilon \in [\varepsilon_{\mathrm{min}},\varepsilon_{\mathrm{max}}]} \mu_E(\varepsilon)\log \mu_E(\varepsilon),
\label{eq:entg-entr-num}
\end{equation}
where $[\varepsilon_{\mathrm{min}},\varepsilon_{\mathrm{max}}]$ is the interval of energies where numerical values were sampled. The behaviour with system size of the classical entanglement $S_{\mathrm{ent}}(N)$ in the homogeneous and in the localized regime, which is the main result of this paper, is represented in Fig.~\ref{fig:entg-entropy-num}. As expected, in the homogeneous positive temperature phase, where the thermodynamics is standard, classical entanglement does not depend on system size: it remains constant upon varying $N$, its value being an offset which can be set to zero without any loss of generality. On the contrary, we find that in the localized phase classical entanglement shows the characteristic behavior
\begin{equation}
S_{\mathrm{ent}}(N) \sim \log(N),
\end{equation}
which is well known to be typical of one-dimensional entangled systems~\cite{CC05,NGR13}. This is to our knowledge the first time such a result has been found for a classical system. In fact, despite the condensate wave-function $z_j$ is a field of quantum origin, in the DNLSE framework it is treated as a classical field obeying Hamiltonian dynamics and is perfectly described by classical ensembles. The presence of a nontrivial behavior of a quantity which is the classical analogue of entanglement entropy, in the context of the present discussion, can be perfectly understood as due to the {\it lack of locality} enforced by the breaking down of the canonical description of the system: a quite new and unexplored regime where classical statistical mechanics seems to capture a phenomenology so far believed to characterize only pure quantum systems~\cite{CC04,CC05,CC09,ECP10}. The logarithmic behavior $S_{\mathrm{ent}} \sim \log(N)$ has also a quite simple interpretation in terms of localization: as long as we identify the number of microstates with the possible locations of the breather on the lattice, which is equal to the number $N$ of lattice sites, it is then natural to say that the entropy of the system is $\log(N)$.\\
\ggcol{\subsection{Classical Entanglement: Large-deviations results}}
\noindent
We show in this section that the numerical result for the classical entanglement in the localized phase of the DNLSE, presented in Fig.~\ref{fig:entg-entropy-num}, is consistent with the analytical predictions which can be drawn from the exact knowledge of the marginal distribution $\mu_E(\varepsilon)$ that we have from~\cite{GILM21a,GILM21b}, which reads as:
\begin{equation}
  \begin{array}{rcl}
    \mu_E(\varepsilon) &=& f(\varepsilon), \qquad\quad ~\varepsilon \ll \Delta E, \\[8pt]
    \mu_E(\varepsilon) &=& g_E(\varepsilon), \qquad\quad \varepsilon \sim \Delta E.
  \end{array}
  \label{eq:equilibrium_distribution}
\end{equation}
with
\begin{equation}
  \begin{array}{rcl}
    f(\varepsilon) &=& \frac{1}{\sqrt{\varepsilon}} \, \exp{\left( - \sqrt{\varepsilon} \right)}, \\[8pt]
    g_E(\varepsilon) &=& \frac{1}{\sqrt{2\pi\sigma^2}N^{3/2}}  
    \, \exp{\left\lbrace -\frac{(\varepsilon -  \Delta E)^2}{2N\sigma^2} \right\rbrace}.
  \end{array}
  \label{eq:marg-th}
\end{equation}
and where $\Delta E = E - E_{\mathrm{th}}$ is the excess energy with respect to the ensemble equivalence threshold $E_{\mathrm{th}} = N e_{\mathrm{th}}$, the energy value corresponding to infinite-temperature line in the phase diagram of Fig.~\ref{fig:pd}. The function $g_E(\varepsilon)$ represents the Gaussian peak associated to the presence of a breather collecting on a single lattice site a macroscopic portion of excess energy $\Delta E$. This Gaussian peak is the well-known {\it condensate bump} characterizing the bimodal shape of $\mu_E(\varepsilon)$ in the localized phase. In Fig.~\ref{fig:marg-th}, which must be compared with the numerical results of Fig.~\ref{fig:marg-loc}, is reported the analytical estimate of $\mu_E(\varepsilon)$ at different values of $N$, where the lattice size influences the position and shape of the secondary peak, see the above Eq.~\ref{eq:marg-th}, since we have $\Delta E = N (e - e_{\mathrm{th}})$. In particular, the distributions of Fig.~\ref{fig:marg-th} are those obtained for $e=4.5 > e_c$. From the explicit analytical knowledge of $\mu_E(\varepsilon)$ in the localized phase, in particular of its explicit dependence on $N$, it is obtained an analytic estimate of the dependence on $N$ of the classical entanglement by exploiting the definition
\begin{equation}
S_{\mathrm{ent}}(N) = - \int_0^{Ne} d \varepsilon~  \mu_E(\varepsilon)  \log\left( \mu_E(\varepsilon)\right)
\end{equation}
The result of the numerical integration of the analytical expression of $\mu_E(\varepsilon)\log \mu_E(\varepsilon)$ is shown in Fig.~\ref{fig:ent-th}, where the asymptotic behavior $S_\mathrm{ent} (N) \sim \log(N)$ can be clearly appreciated. 
This result on classical entanglement suggests a close analogy between the localized phase of the DNLSE and the properties associated to Many-Body Localization (MBL) in genuine quantum interacting systems \cite{NRHA14,L16,DAEB19,SLSVZ25}. In the DNLSE, it is well known that, apart from the condensate total mass and energy, there are no exact conservation laws. An issue that fully deserves a deeper investigation, in the perspective of highlighting the analogies between this system and MBL, is therefore the search of an extensive number of quasi-conserved quantities also in the DNLSE, something which is here left for future investigations.
\newpage

\ggcol{\section{Pseudo-localized regime}}
\label{sec:pseudolocalized}

This section is dedicated to the study of the participation ratio and the microcanonical temperature for energies of the initial condition in the interval $e_{\mathrm{th}} < e < e_c$, where the localized phase is metastable. 
\subsection{Participation ratio} 
\begin{figure}[H]
\centering
\includegraphics[width=0.75\columnwidth]{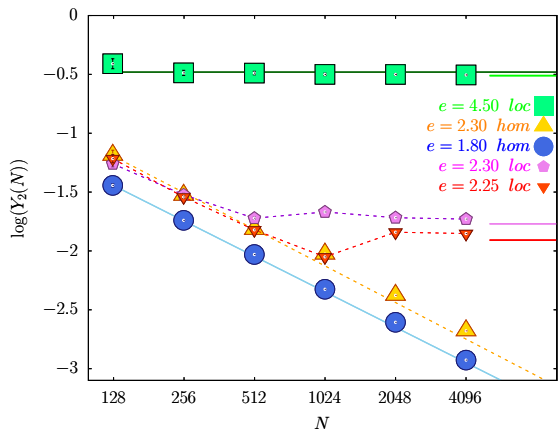}
  \caption{Finite-size scaling of the participation ratio $Y_2(N)$ of the DNLSE for $a=1$ and different energy densities $e$. \emph{Green squares}: $e=4.5$, localized initial conditions ($loc$); \emph{Purple penthagons}: $e=2.3$, localized initial conditions; \emph{red downward triangles}: $e=2.25$, localized initial conditions; \emph{Yellow triangles}: $e=2.3$, homogeneous initial conditions; \emph{Blue circles}: $e=1.8$, homogeneous initial conditions. Statistical expectations are computed as time averages in the time interval $10^5$-$10^6$. Error bars are the standard deviation of fluctuations, sampled at intervals $\Delta t =10$, in the measurement window.}
  \label{fig:part_ratio}
\end{figure}

In the regime $e_{\mathrm{th}}<e<e_c$ we started dynamical simulations from either localized or homogeneous initial conditions. The corresponding data are reported in Fig.~\ref{fig:part_ratio}, where for comparison also the data points already discussed in the main text are shown with reference to Fig.~\ref{fig:part_rat_temp_observables} therein and relative to the homogeneous and localized phase. In Fig.~\ref{fig:part_ratio} data points obtained from localized initial conditions, indicated as $loc$, are purple pentagons and red downward triangles, while those related to homogeneuos initial conditions are yellow triangles. The result is striking, in the sense that it really reveals the presence of two competing equilibrium phases. Starting from a localized state we find that even at moderately large $N$ the participation ratio $Y_2(N)$ relaxes to the equilibrium value of localized phase at $N=\infty$, that is $Y_2 = (e-e_{\mathrm{th}})^2/e^2$, marked by the corresponding horizontal line in the figure. We note that for all sizes $N$ considered in this intermediate regime, we have $e<e_c(N)$, so that the relaxation of $Y_2(N)$ to the value $Y_2(N=\infty)$ is an effect of the choice of initial conditions. In fact, at the very same energies, consider, for instance, the case $e=2.3$ in the figure, by initializing the Hamiltonian dynamics from a homogeneous initial condition, the system remains there, exhibiting a perfect homogeneous-phase scaling of the participation ratio, $Y_2(N) \sim N^{-1}$. Elsewhere, this same regime was found to be characterized by persistent multibreather states~\cite{Iubini2013_NJP} and by a nontrivial chaotic behavior~\cite{Flach2018_PRL,IP21}.
Homogeneous metastable states of the DNLSE were recently studied in~\cite{IP25}.
\\\\
The sensible dependence on initial conditions in the regime $e_{\mathrm{th}} < e < e_c$ reveals the competing phase scenario typical of first-order like transitions. Even when the system is initialized in the metastable state, relaxation can be arbitrarily long, with a characteristic time scale growing exponentially fast with $N$~\cite{PRL_DNLSE}.
Concerning the dependence on $N$ of $Y_2(N)$ for energies in the intermediate regime $e_{\mathrm{th}} < e < e_c$, let us further comment Fig.~\ref{fig:part_ratio} by noticing that the crossover of $Y_2(N)$ from the $1/N$ scaling to a constant value in the intermediate region is consistent, in particular for $e=2.25$ (red downwards triangles), with the same non-monotonic dependence on $N$ observed in~\cite{GIP21,GIP24}. 
\newpage
\newpage
\subsection{Negative temperature} 
\begin{figure}[H]
\centering
  \includegraphics[width=0.75\columnwidth]{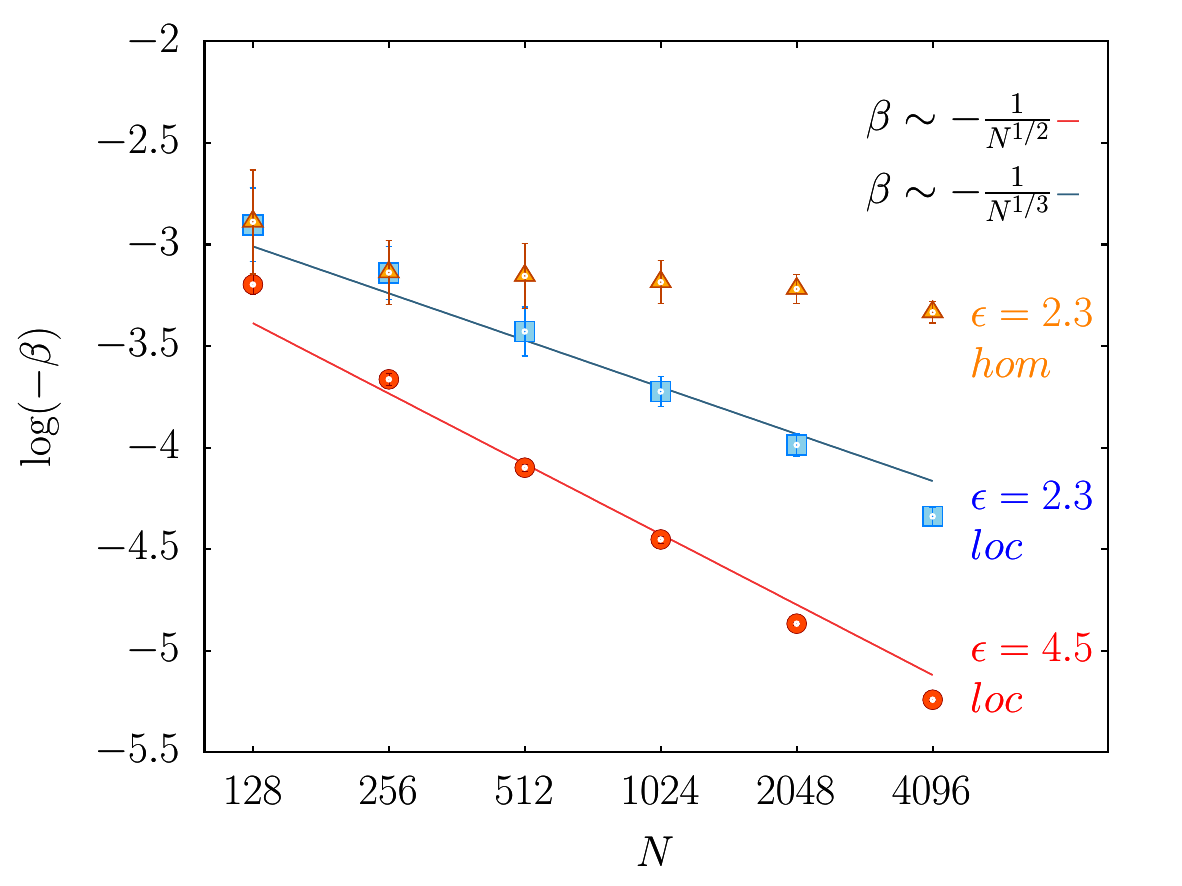}
  \caption{Finite-size scaling of minus the inverse temperature, $-\beta$, vs. $N$ in the metastable and stable localized phases. The system has been initialized with an artificial breather added to an infinite temperature background for the blue and red curves, and with a phase update algorithm for the orange curve. All systems are taken with $a = 1$, while $e = 4.5$ corresponds to the localized phase (red circles) and $e = 2.3$ to the metastable phase (blue squares and orange triangles). The running time of the simulations is $\Delta \tau = 1.5 \times 10^3$ , and the microcanonical temperature has been measured globally across all the sites of the system. The uncertainties have been evaluated as the standard deviation of the temperature, measured at time intervals $d\tau = 10$. The red and blue lines correspond to the theoretical exponents.}
  \label{fig:temp}
\end{figure}

For what concerns the finite-size scaling of temperature in the pseudo-localized regime, $e_{\mathrm{th}}< e <e_c$, we find that the agreement with the analytical predictions of~\cite{GILM21a,GILM21b} depends on the choice of initial conditions. As done for the participation ratio, in Fig.~\ref{fig:temp} we report both the data already shown in Fig.~\ref{fig:part_rat_temp_observables} of the main text, representing the negative temperature of the localized phase, and the data relative to the pseudo-localized phase. In this latter phase we find that for localized initial conditions the dependence of temperature on the system size $N$ seems to better agree with the analytical estimate $\beta \sim - N^{-1/3}$ given in~\cite{GILM21a} for the matching regime, while for homogeneous initial conditions, it deviates more, in particular, $\beta$ seems to decrease with $N$ sensibly slower than $N^{-1/3}$. We consider that the overall agreement between our numerical data on negative temperature in the pseudo-localized phase and analytical predictions is fairly good: in fact, it must be recalled that the behavior $\beta \sim -N^{-1/3}$ is the one typical of the matching regime, that is energies in the neighborhood of $e_c$, while the energy considered here, $e=2.3$, lies strictly below. \\
Furthermore, the dependence of the $\beta$-scaling with $N$ on the choice of initial conditions can also be intuitively explained in terms of the first-order like competing phases scenario, where a remarkably different behavior can be found whether the system is initialized in the stable or metastable phase. And from this respect, neither the localized nor the homogeneous initial conditions represent the true nature of the pseudo-localized phase, which is also characterized by large inhomogeneities in energy distribution across the chain. The difference between the pseudo-localized phase and the fully localized one is that in the former the amount of energy $\varepsilon_j$ concentrated on a given site $j$ can still grow with $N$, but it cannot be proportional to it. 
\newpage

\section{Conclusions}
\label{sec:conclusions}
In this work we have investigated numerically the thermodynamic properties of the localization transition in the Discrete Non-Linear Schr\"odinger Equation (DNLSE). Our results, which confirm the large-deviation analytical estimates of~\cite{GILM21a,GILM21b}, shed light on the controversial nature of the high energy localized phase of the DNLSE, which is now well established as an equilibrium phase in the microcanonical ensemble. We have then clarified how the correct definition of temperature in this phase can only be as a global observable in the microcanonical ensemble. Hamiltonian dynamics gave us validation of the equilibrium finite-size scaling predictions for the participation ratio and the negative temperature contained in~\cite{GILM21a,GILM21b}. On the top of that, our most noticeable result is the behaviour of classical entanglement in the localized phase, which grows as $S_{\mathrm{ent}}(N) \sim \log(N)$. This result sheds light on the subtle and unexpected connection between classical and quantum systems. This is the first example of a field with classical statistics that exhibits a non-trivial behaviour of classical entanglement, a quantity that we have defined in complete analogy with entanglement entropy for quantum systems. This fact is related to the lack of statistical ensembles in the high-energy phase. In this phase the global microcanonical constraint on energy cannot be released without having the disappearance of the localized phase itself. We have therefore ascertained that the localized phase of the DNLSE cannot be interpreted simply as a breather superimposed to an infinite temperature background~\cite{Rumpf2004,rumpf09}: the whole system is entangled so that temperature makes sense as a physical observable only globally.

\section*{Acknowledgements}
We thank for many interesting discussions D. Lucente, V. Ros and L. Salasnich. G.G. thanks the Physics Department of Florence for kind hospitality during some stages in the preparation of this work and acknowledges support from the project MIUR-PRIN2022, “Emergent Dynamical Patterns of Disordered Systems with Applications to Natural Communities”, code 2022WPHMXK. S.I. acknowledges support from the MUR PRIN2022 project ``Breakdown of ergodicity in classical and quantum many-body systems'' (BECQuMB) Grant No. 20222BHC9Z.
\newpage
\appendix

\section{Non-locality of temperature} 
\label{supp:T-nonlocality}
\begin{figure}[H]
\centering
  \includegraphics[width=0.75\columnwidth]{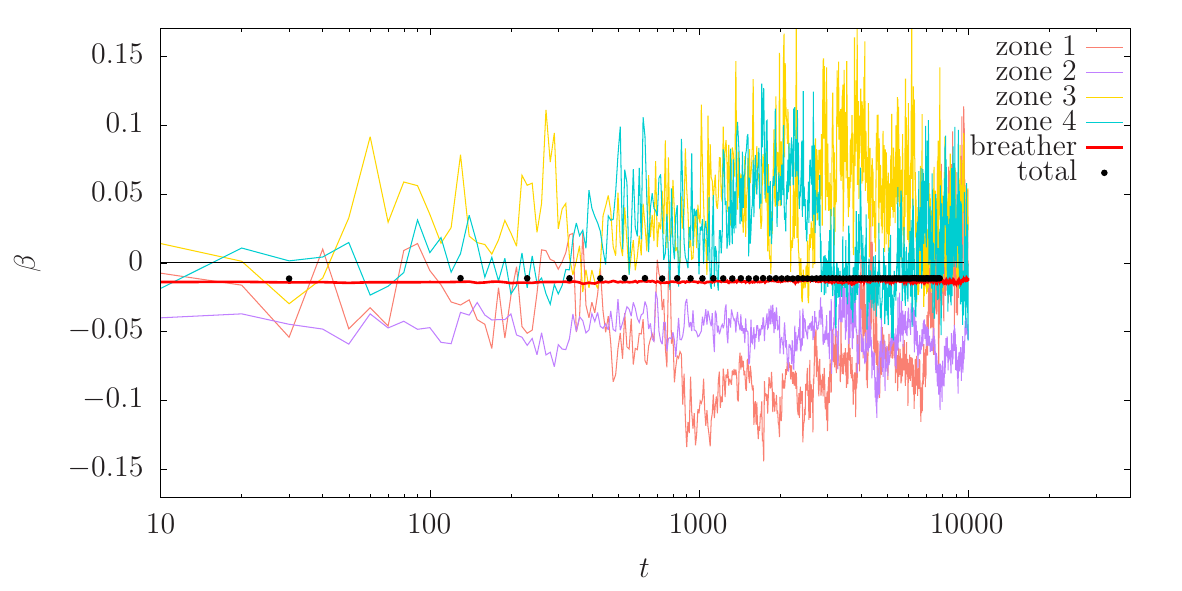}
  \caption{Local measurements of the inverse microcanonical
    temperature $\beta(t)=\mF(\vec{x}(t)) $ as a function of time $t$ for a
    lattice with $N=1024$ points, $a=1$ and energy $e=4.5$. {\it
      Continuous lines}: local value of $\beta$ measured for generic
    chunks of $M=100$ with no breather inside; {\it Red triangles}:
    local value of $\beta(t)$ measured for the chunk of $M=100$ with the
    breather inside; {\it Black dots}: global measurement of $\beta$
    computed over all lattice sites.}
\label{fig:local-temperature}
\end{figure}

As outlined in the main text in Sec.~\ref{sec:therm-obs}, the measure
of a negative microcanonical temperature is a distinguishing feature
of the localized phase: here we wish to present the evidence that it
is precisely from the measure of temperature that we have a strong
hint of the intrinsically non-local nature of the localized phase of
the DNLSE and of its non-additive nature. In order to show this we
performed the measure of $\beta_{\mathrm{micro}}
=\langle\mathcal{F}\{z_j\} \rangle_E$ by computing the function
$\mathcal{F}$ only on sub-portions of the DNLSE lattice. The result is
simple and at the same time striking: even if applying the correct
microcanonical definition, any {\it local} measure of temperature
yields a wrong result. Only if it happens that, by chance, the lattice
portion containing the macroscopic breather is considered, the local
measure of the temperature agrees with the global one. But in the
thermodynamic limit the probability to end up in the system portion
containing the breather is zero, so that it can be stated that, for
any practical purpose, any local measurement of temperature is
wrong. \ggcol{Let us stress that this statement of the inconsistency
  of local measurements of temperature applies only to microcanonical
  equilibrium and to the special {\it ``non-separatbility''} of the
  system emerging in this phase: if the system is driven out of
  equilibrium, as for instance is shown in~\cite{e19090445}, there might well be islands of local
  equilibrium, but the whole description in terms of the global
  microcanonical equilibrium breaks down, with all its more subtle
  consequencies and implications. Very roughly speaking, stationary
  non-equilibrium is a completely different story.} Data for local
measures of microcanonical temperature in the DNLSE are shown in
Fig.~\ref{fig:local-temperature}, in particular the temperature
measured for 5 different portions of a $N=4096$ sites DNLSE lattice,
each portion being made of $M=128$ sites. In
Fig.~\ref{fig:local-temperature} coloured continuous lines represent
the temperatures of different lattice chunks, which are compared with
the global temperature (black bullets). Data are sampled at $e=4.5
>e_c(N)$: it can be clearly seen that in the localized phase local
measures of the temperature do not agree with the global one. On the
contrary, by doing the same comparison between global and local
measures of temperature in the homogeneous ``thermal'' phase of the
system (data for this case are not shown), $e<e_{\mathrm{th}}$, we
have found agreement between the two measures of temperature, as can
be expected from the equivalence between the canonical and
microcanonical ensembles.
These results for temperature in the localized regime, $e>e_c(N)$,
confirm that for the DNLSE in this particular phase the temperature
$T$ cannot be traced back to a local quantity related to the average
kinetic energy. This fact is not only a consequence of the lack of a
standard kinetic-energy term in the DNLSE Hamiltonian, but is due to
the nature of the localized phase, where $T$ is a function of all
degrees of freedom (see, \cite{rugh97,franzosi11}).

\section{Ergodicity breaking}
\label{supp:dynamics}
\begin{figure}[H]
\centering
  \includegraphics[width=0.75\columnwidth]{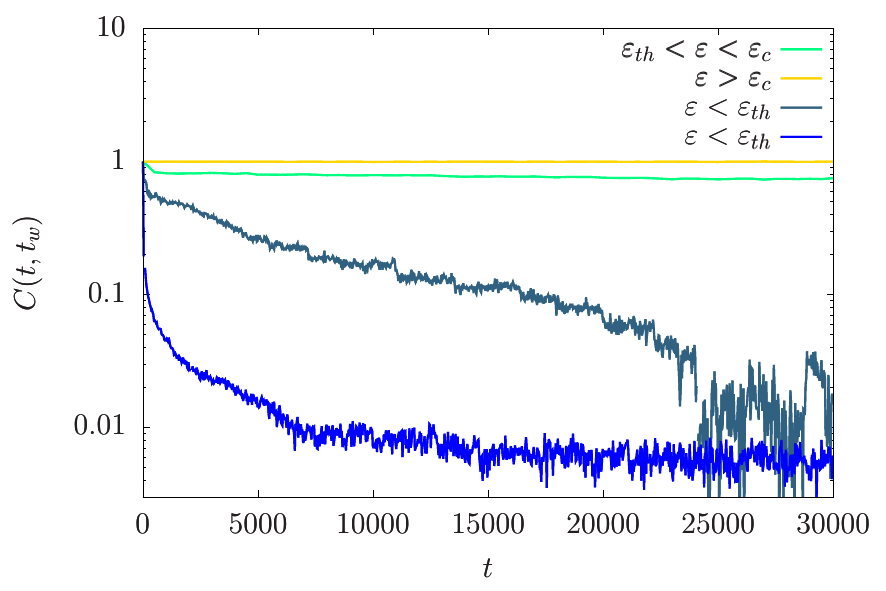}
  \caption{Time correlation function $C(t,t_w)$ for a $N=1024$ DNLSE
 lattice at different values of energy and for different initial conditions. Energy values: $e=4.5>e_c$, negative temperature localized phase (yellow line); $e_{\mathrm{th}} < e =2.3 < e_c$, negative temperature pseudo-localized phase (green line); $e=1.8<e_{\mathrm{th}}$, positive temperature homogeneous phase. For the homogeneous phase, $e < e_{\mathrm{th}}$, we have considered both localized (gray line) and homogeneous (blue line) initial conditions. The curves have been obtained by averaging over independent initial conditions (10 for $e>e_c$, 100 for $e_{\mathrm{th}} < e < e_c$, 50 for $e < e_{\mathrm{th}}$ localized initial conditions, 500 for $e < e_{\mathrm{th}}$ homogeneous initial conditions).}
  \label{fig:Correlation_small}
\end{figure}

Despite all results presented in the main text of this paper are focused on the DNLSE thermodynamics in the localized phase, dynamical properties of this system are also an extremely interesting chapter on its own. To maintain the discussion of this work self-contained we have decided to present just a brief account of the dynamical properties of the system in this section, leaving further investigations for a future work, as for instance the study of the similarities between the DNLSE localization transition and the ergodicity breaking transitions in disordered systems. It must be also acknowledged that the study of the DNLSE slow dynamics, apart from comparisons with the dynamics of glassy systems, has already attracted a lot of interest, with many interesting works already dedicated to this specific subject~\cite{Flach2018_PRL,PRL_DNLSE}.\\
For what concerns the present discussion we just want to present a hint that, consistently with the most naive expectation, localization corresponds to a form of ergodicity breaking. The evidence of this is provided by the behavior of time correlation functions $C(t,t_w)$, for which numerical data are shown in Fig.~\ref{fig:Correlation_small}. Since the variable of interest throughout the whole discussion in this work is the non-linear contribution to local energy of the condensate wave-function, $\varepsilon_i(t) = |z_i(t)|^4$, we have considered the following definition for the two-time correlation function:
\begin{equation}
C(t,t_w) = \frac{1}{\mZ(t_w)} \sum_{i=1}^N \varepsilon_i(t) \varepsilon_i(t_w), 
\end{equation}
where $\mZ(t_w) = \sum_{i=1}^N \varepsilon_i^2(t_w)$ provides the correct normalization. The naive expectation that the pseudo-localized, $e_{\mathrm{th}} < e < e_c$, and localized, $e > e_c$, regimes are also ergodicity-broken phases seems confirmed from a first analysis of the correlation function $C(t,t_w)$, in which we have kept for full generality the dependence on both initial, $t_w$, and final, $t$, time. We have denoted the initial time $t_w$ in order to stress the analogy with the waiting time of glass-forming systems. When the system is initialized with a finite fraction of the total energy localized in a single breather, data in Fig.~\ref{fig:Correlation_small} show that, up to $t=3.0\cdot 10^4$ correlations either never decay in the localized phase, $e=4.5>e_c$, or decay to a plateau at $C^* \sim 0.75$ in the pseudo-localized one, $e=2.3$. We also studied how initial data evolve at energies $e < e_{\mathrm{th}}$ where there is no localization, temperature is positive and equivalence between canonical and microcanonical ensemble holds.\\
\noindent
 In particular, we have studied how correlations decay for an initial condition at $e=1.8 < e_{\mathrm{th}}$: we find that initial conditions with a finite fraction of energy on a single lattice site decay to $1\%$ of the initial value within times $t = 3.0\cdot 10^4$, whereas homogeneous initial conditions at the same energy decay exponentially fast. These data do not allow us to determine the specific type of ergodicity breaking transition observed or to characterized the aging dynamics at work. They just provide the evidence that localization is accompanied by a dynamical arrest, encouraging to further investigations in this direction.

\newpage
\section{Microcanonical Temperature}
\label{supp:F}
An operative definition of microcanonical temperature $T^{-1}=\partial S / \partial E$ is obtained following the results in~\cite{franzosi11}, where the cartesian real coordinates $x_{2j}=q_j\equiv\sqrt{2}~\mathrm{Im}(z_j)$, $x_{2j+1}=p_j\equiv \sqrt{2}~\mathrm{Re}(z_j)$ are introduced, so that
\begin{equation}
\vec{x} = \lbrace p_1,q_1,p_2,q_2,\ldots,p_N,q_N\rbrace.
\end{equation}
In this representation the energy and total mass appearing in Eq.(\ref{norm.1}) and Eq.~(\ref{energy}) can be written as
\begin{eqnarray}
\mH&=&\sum_{j=1}^N \frac{1}{4} (p_j^2 + q_j^2)^2 + \sum_{j=1}^N (p_j p_{j+1} + q_j q_{j+1})\\
\mA&=&\frac{1}{2} \sum_{j=1}^N (p_j^2 + q_j^2),
\end{eqnarray} 
 or, equivalently, as
\begin{eqnarray}
\mH &=&\sum_{j=1}^N \frac{1}{4} (x_{2j}^2 + x_{2j+1}^2)^2 + \sum_{j=1}^N (x_{2j+1}x_{2j+3} + x_{2j} x_{2j+2}) \nonumber \\
\mA &=&\frac{1}{2} \sum_{j=1}^{2N} x_j^2.
\end{eqnarray} 
In terms of these variables one can then write the global function $\mF(\vec{x})$ such that $\beta = T^{-1} = \langle \mF(\vec{x}) \rangle_E$ as
\begin{equation}
\label{eq:Tmicro}
\mF(\vec{x})=\frac{W(\vec{x})\|\vec \xi\|}{\vec\nabla \mH\cdot\vec\xi}\ 
\vec\nabla\cdot\left(\frac{\vec\xi}{W(\vec{x})\|\vec\xi\|} \right)
\end{equation}
where the vector $\vec{\xi}$ and the function $W(\vec{x})$ are defined as
\begin{eqnarray}
\vec{\xi} &=& \frac{\vec{\nabla} \mH}{\|\vec{\nabla} \mH\|} - \frac{(\vec{\nabla} \mH \cdot \vec{\nabla} \mA) \vec{\nabla} \mA}{\|\vec{\nabla} \mH\| \|\vec{\nabla} \mA\|^2} \\[8pt]
W(\vec{x}) &=& \sqrt{\sum_{j=1}^{2N} \sum_{k=j+1}^{2N} \left[ \frac{\partial \mH}{\partial x_j} \frac{\partial \mA}{\partial x_k} - \frac{\partial \mH}{\partial x_k} \frac{\partial \mA}{\partial x_j} \right]^2} \, . \nonumber
\end{eqnarray}
Notice that the resulting definition Eq.~(\ref{eq:Tmicro}) is nonlocal in space, as it involves the contribution of all degrees of freedom.
\newpage
\section*{References}
\bibliographystyle{iopart-num}
\bibliography{biblio}
\end{document}